\documentstyle[12pt,psfig]{article}
\begin{document}

\hfill DUKE-CGTP-2000-13

\hfill hep-th/0008154

\vspace{1.5in}

\begin{center}

{\large\bf Discrete Torsion }

\vspace{1in}

Eric Sharpe \\
Department of Physics \\
Box 90305 \\
Duke University \\
Durham, NC  27708 \\
{\tt ersharpe@cgtp.duke.edu} \\

 $\,$

\end{center}

In this article we explain discrete torsion.
Put simply, discrete torsion is the choice of orbifold group
action on the $B$ field.  We derive the classification
$H^2(\Gamma, U(1))$, we derive the twisted sector phases
appearing in string loop partition functions, we derive
M.~Douglas's description of discrete torsion for D-branes
in terms of a projective representation of the orbifold group,
and we outline how the results of Vafa-Witten fit into this
framework.  In addition, we observe that additional degrees of freedom
(known as shift orbifolds)
appear in describing orbifold group actions on $B$ fields,
in addition to those classified by $H^2(\Gamma, U(1))$, and explain
how these degrees of freedom appear in terms of twisted sector
contributions to partition functions and in terms of orbifold
actions on D-brane worldvolumes.  
This paper represents a
technically-simplified version of prior papers by the author
on discrete torsion.
We repeat here technically simplified versions of results from
those papers, and have included some new material.

\begin{flushleft}
August 2000
\end{flushleft}

\newpage

\tableofcontents

\newpage

\section{Introduction}

Historically discrete torsion has been a rather mysterious
aspect of string theory.  Discrete torsion was originally
discovered \cite{vafa1} as an ambiguity in the choice of
phases to assign to twisted sectors of string orbifold partition
functions.  Although other work has been done on the subject
(see, for example, \cite{vafaed,doug1,doug2}), no work done prior to
\cite{dt1,dt2} 
has succeeded in giving any sort of 
genuinely deep understanding
of discrete torsion.  In fact, discrete torsion has sometimes been
referred to has an inherently stringy degree of freedom, without
any geometric analogue.

In this paper (a followup to \cite{dt1,dt2}) we shall describe a purely
mathematical way of understanding discrete torsion,
and will show explicitly how our description gives rise
to Vafa's phases in twisted sector contributions to partition functions,
to projective representations of orbifold group actions on D-branes,
and to other physical manifestations of discrete torsion.

The description of discrete torsion we present here is the same
as that we previously presented in \cite{dt1,dt2}.  
This paper differs in that we
have vastly reduced the level of technical complication that
was present in \cite{dt1,dt2}, we explicitly work out the details
of some computations merely referred to in \cite{dt1,dt2}, and we also
work out some new results not present in \cite{dt1,dt2}, such as
a derivation of the projectivized group actions used by
\cite{doug1,doug2} to describe orbifold group actions on D-branes
with discrete torsion.

What is discrete torsion?
In a nutshell,
\begin{center}
{\it Discrete torsion is the choice of orbifold group action on the $B$
field.}
\end{center}

More generally, in any theory possessing fields with gauge invariances,
defining the orbifold group action on the base space does
not suffice to define the orbifold group action on the fields of the theory
-- one can combine the action of the orbifold group with gauge
transformations to get new, distinct, actions of the orbifold group.
For $U(1)$ gauge fields, this gives rise to orbifold $U(1)$ Wilson lines.
For $B$ fields, this gives rise to discrete torsion, as we shall
work through in detail in this paper.
Now, string theory has other fields with gauge invariances -- for example,
in eleven-dimensional supergravity there is a three-form potential
with a gauge invariance analogous to that of the $B$ field, and so
one expects to have orbifold degrees of freedom associated to that
field also.  We shall discuss analogues of discrete torsion for the
other tensor-field potentials of string theory in \cite{cdt}. 
In another upcoming paper \cite{hdt} we shall discuss
discrete torsion in perturbative heterotic strings.

It should be emphasized that we have presented a first-principles
{\it explanation} of discrete torsion in \cite{dt1,dt2} and in
this paper -- not some observations on discrete torsion, not
some calculations related to discrete torsion, but an
explanation.  It should also be emphasized that this explanation
is purely mathematical in nature -- although we can certainly 
check physical consequences, at its heart discrete torsion
is a natural mathematical consequence of having 
$B$ fields.  Discrete torsion has nothing at all to do with
string theory {\it per se}, and is not ``inherently stringy'' in any sense.

We begin in section~\ref{oWlrev} with a discussion of orbifold
$U(1)$ Wilson lines -- i.e., a discussion of counting orbifold
group actions on $U(1)$ gauge fields.  Although the technical
details for orbifold group actions on $B$ fields are considerably
more complicated and subtle, the basic principles are the same for 
$B$ fields as for $U(1)$ gauge fields, and our treatment of $B$ fields
parallels our treatment of $U(1)$ gauge fields.  
In particular, section~\ref{oWlrev} 
provides a simplified context
in which to see the main ideas at work.

Next in section~\ref{derivh2} we study orbifold group actions
on $B$ fields, and explain how the group cohomology group
$H^2(\Gamma, U(1))$ arises.  In addition to $H^2(\Gamma, U(1))$,
we also find additional orbifold group actions; we explain why, in hindsight,
such additional actions should be expected.
The general methods used are the
same as for studying orbifold group actions on $U(1)$ gauge
fields, although the technical details are more complicated.

In section~\ref{twphase} we derive the twisted sector phases originally
described in \cite{vafa1}.  We calculate the phases at one-loop,
and also check factorization at higher loops (an explicit calculation
is done at two-loops, from which the general result is obvious).

In section~\ref{derivdoug} we derive M. Douglas's description
\cite{doug1,doug2} of discrete torsion for D-branes, as a projective
representation of the orbifold group.

In section~\ref{derivvw} we outline how results of C. Vafa and
E. Witten \cite{vafaed} on the interplay of discrete torsion and
Calabi-Yau moduli are naturally understood in the context we have
presented.  This discussion frequently refers to reflexive sheaves
on singular varieties, and as such ideas are not frequently
used in the physics literature, we have included an appendix on the subject.

Finally in section~\ref{locorb} we very briefly mention local
orbifold degrees of freedom, as distinct from global orbifold degrees
of freedom, a topic usually neglected in physics treatments of
orbifolds.

Readers who wish to study this paper in detail are 
encouraged to first work through section~\ref{oWlrev} on
orbifold $U(1)$ Wilson lines, despite the fact that it might
not sound wholly relevant.  Our approach to thinking about
$B$ fields and discrete torsion is very closely related to
orbifold $U(1)$ Wilson lines, and the mathematical techniques
we shall use to study orbifold group actions on $B$ fields
are precise analogues of those used
to classify orbifold $U(1)$ Wilson lines.  Mathematicians should
note that although these mathematical
techniques are a staple of the relevant part of the mathematics
literature, they are not widely used in the physics literature.

\section{Review of orbifold $U(1)$ Wilson lines}   \label{oWlrev}

In this paper we shall describe discrete torsion as a choice
of orbifold group action on $B$ fields.  In order to understand
how this works, however, we shall first present orbifold
$U(1)$ Wilson lines.  We will work through this simpler case
in detail because it is an exact model for our approach to
understanding orbifold group actions on $B$ fields -- the details
are much more subtle, but the basic ideas are the same as for
orbifold $U(1)$ Wilson lines.

In particular, 
our approach to understanding discrete torsion is closely
modelled on understanding orbifold $U(1)$ Wilson lines mathematically,
involving techniques not commonly used in the physics literature.
In order to understand later sections of this paper, therefore,
it is important to first get a solid handle on the basic ideas,
in a context in which the details are easy to work out.

Orbifold $U(1)$ Wilson lines are precisely a choice of orbifold group
action on $U(1)$ gauge fields.  We shall first work out a description
of elements of the set of orbifold group actions on $U(1)$ gauge fields,
then we shall argue that any two elements of the set differ
by an element of the group $H^1(\Gamma, U(1))$.

Our analysis does not rely on the orbifold group $\Gamma$
acting freely -- whether $\Gamma$ has fixed points is entirely
irrelevant for our analysis.  Similarly, whether $\Gamma$
is abelian is equally irrelevant.

It should also be mentioned that our analysis of orbifold
actions on $U(1)$ gauge fields is not new, but is rather
quite standard in the mathematics literature.  The earliest
reference of which we are aware is \cite[section 1.13]{kostant}.

For simplicity, we shall work at the level of transition functions.
A line bundle with connection (i.e., a set of local $U(1)$ gauge fields)
can be described by pairs $( A^{\alpha}, g_{\alpha \beta})$
where $A^{\alpha}$ is a vector field on an element $U_{\alpha}$
of an open cover, and $g_{\alpha \beta}$ are transition functions.
These are related by
\begin{equation}     \label{atog1}
A^{\alpha} \: - \: A^{\beta} \: = \: d \log g_{\alpha \beta}
\end{equation}

\subsection{Orbifold group actions on vector fields}

In this subsection we shall study the set of orbifold
group actions on vector fields, in terms of transition functions
and data defined on local coordinate patches.  To describe
the action of the orbifold group on such data, one relates
pullbacks $g^* g_{\alpha \beta}$, for example, to the original
transition functions $g_{\alpha \beta}$.  
So, in this subsection we shall work out
relationships between pullbacks of transition functions $g^* g_{\alpha \beta}$
and gauge
fields $g^* A^{\alpha}$, and original transitions functions $g_{\alpha
\beta}$ and gauge fields $A^{\alpha}$.

For convenience, we shall choose an open cover that is well-behaved
with respect to the action of the orbifold group.
Specifically, let $\{ U_{\alpha} \}$ be a ``good invariant'' cover,
meaning that each $g \in \Gamma$ maps each $U_{\alpha}$ back
into itself (i.e., $g: U_{\alpha} \rightarrow U_{\alpha}$),
and each $U_{\alpha}$ is a union of disjoint contractible open sets.
Such a cover is not a good cover, because the elements $U_{\alpha}$
will not be contractible in general, but is the next best thing,
and suffices for our purposes.

To begin, we need to demand
that the bundle is isomorphic to itself
under pullback by group elements, i.e., that the bundle itself is
``symmetric'' with respect to the group action.
Given this constraint, we will derive the form of the pullback
of the $U(1)$ gauge field from self-consistency.
At the level of transition functions, this is the statement
\begin{equation}
g^* g_{\alpha \beta} \: = \: \left( h^g_{\alpha} \right) \,
\left( g_{\alpha \beta} \right) \, \left( h^g_{\beta} \right)^{-1}
\end{equation}
for some \v{C}ech cochains $h^g_{\alpha}$.  (Such cochains
define an isomorphism from the bundle itself to its pullback by
$g$.)  If such a statement were not true, one could not even
begin to define an orbifold group action, as it would mean
that in no sense is the bundle well-behaved with respect to the
orbifold group.

Next, we need to determine how $h^{g_1 g_2}_{\alpha}$ is related to
$h^{g_1}_{\alpha}$ and $h^{g_2}_{\alpha}$.
We can find such a constraint by expanding $(g_1 g_2)^* g_{\alpha \beta}$
in two different ways:
\begin{eqnarray*}
(g_1 g_2)^* g_{\alpha \beta} & = & \left( h^{g_1 g_2}_{\alpha} \right)
\, \left( g_{\alpha \beta} \right) \, 
\left( h^{g_1 g_2}_{\beta} \right)^{-1} \\
\mbox{ also } & = & g_2^* \left[ \,
\left( h^{g_1}_{\alpha} \right) \, \left( g_{\alpha \beta} \right) \,
\left( h^{g_1}_{\beta} \right)^{-1} \, \right] \\
 & = & \left( g_2^* h^{g_1}_{\alpha} \right) \,
\left( h^{g_2}_{\alpha} \right) \,
\left( g_{\alpha \beta} \right) \,
\left( h^{g_2}_{\beta} \right)^{-1} \,
\left( g_2^* h^{g_1}_{\beta} \right)^{-1} 
\end{eqnarray*}
From self-consistency, we see that it is natural to demand
\begin{equation}  \label{h12}
h^{g_1 g_2}_{\alpha} \: = \: \left( g_2^* h^{g_1}_{\alpha} \right) \,
\left( h^{g_2}_{\alpha} \right)
\end{equation}

We should take a moment to comment on this ``derivation.''
All we really know is that the \v{C}ech coboundary of the $h^g_{\alpha}$
satisfies an equation of the form above -- strictly speaking, it is
not quite true that equation~(\ref{h12}) necessarily follows.
However, we are looking for constraints of the general form of
equation~(\ref{h12}), and from the previous algebra 
equation~(\ref{h12}) emerges quite naturally.  If the reader
prefers, it might be slightly more fair to say that we are
using self-consistency to bootstrap an ansatz.  
We shall use similar methods many more times, both here
and in working out orbifold group actions on $B$ fields. 

In addition, in this special case, there is an additional concern
that the reader might have. 
In principle, we could multiply
one side of equation~(\ref{h12}) by an $\alpha$-independent phase,
and the result would still be consistent.  We are
implicitly imposing a slightly stronger constraint than 
strictly necessary -- namely, that the orbifold group action
be honestly represented.  This choice is precisely the choice that
leads to the calculation of orbifold Wilson lines, for example.

In passing, note that equation~(\ref{h12}) can be viewed as 
the statement that the map
from the orbifold group $\Gamma$ to bundle isomorphisms
(as defined by \v{C}ech cohains $h^g_{\alpha}$) is a group
homomorphism.

At this point we have derived the form of an equivariant structure
on the principal $U(1)$ bundle itself, but have not mentioned
the connection on the bundle.
Without loss of generality, define $\varphi^{\alpha}_g$ by
\begin{displaymath}
g^* A^{\alpha} \: = \: A^{\alpha} \: + \:
\varphi^{\alpha}_g
\end{displaymath}
Certainly, regardless of $g^* A^{\alpha}$, we can write this for
some $\varphi^{\alpha}_g$, so all we have done is define
$\varphi^{\alpha}_g$, not place any sort of constraint on the
connection $A^{\alpha}$.

By expanding $(g_1 g_2)^* A^{\alpha}$ in two different ways,
one quickly finds
\begin{displaymath}
\varphi^{\alpha}_{g_1 g_2} \: = \: \varphi^{\alpha}_{g_2}
\: + \: g_2^* \varphi^{\alpha}_{g_1}
\end{displaymath}

By pulling back equation~(\ref{atog1}) by $g$,
one finds
\begin{equation}
\varphi^{\alpha}_g \: = \: d \log h^g_{\alpha}
\end{equation}

So far we have worked out how to describe the action of an
orbifold group $\Gamma$ on a principal $U(1)$ bundle with
connection (a set of $U(1)$ gauge fields, if the reader prefers).
To summarize our results so far, we can describe this action
at the level of transition functions as
\begin{eqnarray*}
g^* A^{\alpha} & = & A^{\alpha} \: + \: d \log h^g_{\alpha} \\
g^* g_{\alpha \beta} & = & \left( h^g_{\alpha} \right) \,
\left( g_{\alpha \beta} \right) \,
\left( h^g_{\beta} \right)^{-1} \\
h^{g_1 g_2}_{\alpha} & = & \left( h^{g_2}_{\alpha} \right) \,
\left( g_2^* h^{g_1}_{\alpha} \right)
\end{eqnarray*}
for some \v{C}ech cochains $h^g_{\alpha}$, which define the orbifold
group action.

We have described the action of the orbifold group in terms
of transition functions and data on local charts, but one can
also describe the same orbifold group action more elegantly
in terms of an action on the total space of a bundle.
See for example \cite{dt1} where this approach is reviewed.

\subsection{Differences between orbifold group actions}

In the previous subsection we described elements of the set of
orbifold group actions on $U(1)$ gauge fields (more properly,
principal $U(1)$ bundles with connection).
We have not used the term set loosely -- in general, there is
no natural way to place a group structure, for example, on this set.
In this section we shall point out that any two such actions
(i.e., any two elements of the set)
differ by a constant gauge transformation, and those constant
gauge transformations lead to $H^1(\Gamma, U(1))$.

Let $h^g_{\alpha}$ be one set of \v{C}ech cochains describing
an orbifold action on a set of $U(1)$ gauge fields,
and let $\overline{h}^g_{\alpha}$ be a 
distinct orbifold action on the same set of $U(1)$ gauge fields.

Define \v{C}ech cohains $\phi^g_{\alpha}$ by
\begin{displaymath}
\phi^g_{\alpha} \: = \: \frac{ h^g_{\alpha} }{
\overline{h}^g_{\alpha} }
\end{displaymath}

From writing $g^* g_{\alpha \beta}$ in terms of the two orbifold actions,
we find a constraint on the $\phi^g_{\alpha}$:
\begin{eqnarray*}
g^* g_{\alpha \beta} & = & \left( h^g_{\alpha} \right) \,
\left( g_{\alpha \beta} \right) \, \left( 
h^g_{\beta} \right)^{-1} \\
\mbox{ also } & = & \left( \overline{h}^g_{\alpha} \right) \,
\left( g_{\alpha \beta} \right) \, 
\left( \overline{h}^g_{\beta} \right)^{-1} 
\end{eqnarray*}
Dividing these two lines we find that
$\phi^g_{\alpha} = \phi^g_{\beta}$ on $U_{\alpha} \cap U_{\beta}$,
i.e., the $\phi^g_{\alpha}$ define a function,
which we shall denote $\phi^g$.

Next, write $g^* A^{\alpha}$ in terms of the two orbifold group actions
to find an additional constraint:
\begin{eqnarray*}
g^* A^{\alpha} & = & A^{\alpha} \: + \: d \log h^g_{\alpha} \\
\mbox{ also } & = & A^{\alpha} \: + \:
d \log \overline{h}^g_{\alpha} 
\end{eqnarray*}
Subtracting these two lines, we find that
\begin{displaymath}
d \log \phi^g \: = \: 0
\end{displaymath}

In other words, $\phi^g$ is a constant function.

Finally, from
\begin{eqnarray*}
h^{g_1 g_2}_{\alpha} & = & \left( h^{g_2}_{\alpha} \right)
\, \left( g_2^* h^{g_1}_{\alpha} \right) \\
\overline{h}^{g_1 g_2}_{\alpha} & = &
\left( \overline{h}^{g_2}_{\alpha} \right) \,
\left( g_2^* \overline{h}^{g_1}_{\alpha} \right)
\end{eqnarray*}
we find that
\begin{equation}
\phi^{g_1 g_2} \: = \: \phi^{g_2} \, g_2^* \phi^{g_1}
\end{equation}

Assuming the covering space of the orbifold is connected,
we see that the $\phi$ define a group homomorphism
$\Gamma \rightarrow U(1)$.

In other words, the difference between any two orbifold group
actions on $U(1)$ gauge fields, on a connected space,
is defined by an element of $H^1(\Gamma, U(1))$.

\subsection{General analysis}

In the previous two subsections we did two things -- we
worked out the structure of an orbifold group action on a set
of $U(1)$ gauge fields, and then we argued that any two orbifold
group actions differ by a (constant) gauge transformation,
defining an element of $H^1(\Gamma, U(1))$.

One point mentioned earlier, and worth emphasizing, is that our
derivation of the group $H^1(\Gamma, U(1))$ did not
rely upon $\Gamma$ being freely acting -- the derivation is the same
regardless of whether or not the action of the orbifold
group $\Gamma$ has fixed points.  The same statement will be
true of our derivation of $H^2(\Gamma, U(1))$ in understanding
orbifold group actions on $B$ fields in the next section --
our derivation holds regardless of whether or not $\Gamma$ acts
freely.

Another point worth mentioning is that we have not assumed
that $\Gamma$ is abelian -- our derivation of $H^1(\Gamma, U(1))$
is the same regardless of whether $\Gamma$ is abelian or nonabelian.

We should also note that we have not assumed that the
principal $U(1)$ bundle on which we have defined the orbifold group
action, is trivial.  If the bundle is nontrivial, then one does
not expect to always be able to define an orbifold group
action, even on the topological bundle, much less on the bundle
with connection.  For example, consider the Hopf fibration of
$S^3$ over $S^2$, viewed as a principal $U(1)$ bundle over $S^2$.
Consider a $U(1)$ acting on the $S^2$ by rotations about some fixed
axis.  A $2 \pi$ rotation of the $S^2$ is the same as the identity
action on $S^2$; however, a $2 \pi$ rotation of the $S^2$ does not lift
to the action of the identity on $S^3$ -- one must rotate the $S^2$ by a
multiple of $4 \pi$ instead, as is discussed in most elementary quantum
mechanics textbooks in the context of spin.
See for example \cite{dt1} for further explanation
of the standard well-known fact that group actions on base spaces do not
always lift to nontrivial bundles.  Assuming that group actions on the
bundle exist, the difference between any two group actions on a
principal $U(1)$ bundle with connection is defined by an element of
$H^1(\Gamma, U(1))$.

\subsection{The set of orbifold group actions is a {\it set}}

It should be emphasized that the set of orbifold group actions
on a principal $U(1)$ bundle with connection, is a {\it set}, 
and in general does not naturally have a group structure.
Often in the physics literature, calculations boil down to
calculating some cohomology group -- by contrast, 
possible orbifold group actions do not (in general) have a
group structure, and certainly can not be understood
in terms of a calculation of a cohomology group.

Now, in special cases, it is possible to put a group
structure on the set of orbifold group actions.
For example, if the principal $U(1)$ bundle is topologically
trivial, then there is a natural notion of a trivial
action -- since the base and the fiber can be globally split,
one could take the orbifold group to act on the base only.
In this special case, we can describe any other orbifold group action
in terms of the trivial action plus an element of $H^1(\Gamma, U(1))$
 -- this is the precise technical meaning of ``combining the
action of the orbifold group with a gauge transformation,''
as is often mentioned in the old string orbifold literature.

In general, however, there will be no natural notion of a
``trivial'' action -- all the orbifold group actions will have
some nontrivial action on the fibers of the bundle, and so the
set of orbifold group actions is no more than a set\footnote{
A set naturally acted on by the group $H^1(\Gamma, U(1))$, but a
set nonetheless.}.  

A simple example of these notions is provided by line bundles
on toric varieties.  Specifying a specific toric divisor
is the same as specifying an action of the algebraic
torus underlying the toric variety on the line bundle
(see \cite{sotvfp} and references therein).
For example, consider ${\bf P}^2$ as a toric variety,
with toric divisors generated by $D_x$, $D_y$, and $D_z$.
Possible toric divisors for a degree 0 bundle on ${\bf P}^2$
include $D_x - D_y$, $2 D_z - D_x - D_y$, and so forth -- countably
many, counted by\footnote{For holomorphic line bundles, rather than
principal $U(1)$ bundles with connection, there is a closely
analogous argument relating different choices of actions of
algebraic groups.  The result is essentially the same,
modulo replacing $U(1)$ with ${\bf C}^{\times}$.} 
$H^1(({\bf C}^{\times})^n, {\bf C}^{\times}) = 
{\bf Z}^n$.  For a degree
0 line bundle, one can put a group structure on the set of
degree 0 divisors -- take the identity to be the toric divisor 0.
Now, consider a degree 1 line bundle.  Possible toric divisors
include $D_x$, $D_y$, $3 D_x - 2D_z$, and so forth -- countably
many, counted by $H^1(({\bf C}^{\times})^n, {\bf C}^{\times}) = 
{\bf Z}^n$.  However,
here there is no natural divisor to associate with the identity
 -- the set does not have a group structure in any natural way.
So, we see here explicitly that in general the set of 
orbifold group actions is only a set.

In special cases, such as when the bundle is topologically
trivial, there is a canonical trivial orbifold group action,
and in such cases we can put a group structure on the set of orbifold
group actions, which becomes the group $H^1(\Gamma, U(1))$.
As luck would have it, such special cases are the only ones ever usually 
considered by physicists, so most of the subtleties of the general
case are omitted from typical physics discussions.
These matters are discussed in more detail in \cite{dt1}.

In the next section we shall perform a closely analogous
computation for $B$ fields.  We shall first study elements of
the set of orbifold group actions on the $B$ fields,
then we shall study how different elements of this set are
related.  The technical details for $B$ fields are much more
complicated, but the basic approach is the same.

\section{Derivation of $H^2(\Gamma, U(1))$}     \label{derivh2}

In this section we shall explain how the group cohomology
group $H^2(\Gamma, U(1))$ appears when describing orbifold group
actions on $B$ fields.  Our methods will closely
mirror standard methods used to study orbifold $U(1)$ Wilson lines, so 
readers are encouraged to study section~\ref{oWlrev} before 
reading this section.  To be brief, we first derive the structure
of elements of the set of orbifold group actions on $B$ fields,
and then study the difference between any two elements of this
set.  The difference is a gauge transformation of $B$ fields,
just as the difference between two orbifold group actions on
a set of $U(1)$ gauge fields is a gauge transformation.

A gauge transformation of a $B$ field is defined by a principle
$U(1)$ bundle with connection -- a set of $U(1)$ gauge fields,
if the reader prefers.  So, to each element $g$ of the orbifold group
$\Gamma$,
the difference between any two orbifold group actions on a set of
$B$ fields is defined by a principal $U(1)$ bundle with connection,
call it $T^g$.  (In fact, the connection on $T^g$ is constrained
to be flat, just as the difference between any two orbifold group
actions on $U(1)$ gauge fields was a constant gauge transformation.)
We must also specify isomorphisms 
$\omega^{g,h}$ between
$T^{gh}$ and\footnote{We will explain the meaning of $\otimes$ for
principal $U(1)$ bundles later in this section.} $T^h \otimes h^* T^g$ 
for each $g, h \in \Gamma$, just as for orbifold Wilson lines,
the gauge transformations $\phi^g$ were constrained to obey
$\phi^{g_1 g_2} = \phi^{g_2} \cdot g_2^* \phi^{g_1}$.

This description is not complete -- there are ``residual gauge invariances,''
specifically, only the isomorphism class of $T^g$ is actually relevant.
To get a precise counting, one must fix this residual gauge invariance.

Elements of $H^2(\Gamma, U(1))$ arise from taking the bundle $T^g$ to
be topologically trivial, with a connection that is gauge-equivalent to
zero.  Using residual gauge invariances, we can replace
$T^g$ with the canonical trivial bundle with identically zero connection,
and the maps $\omega^{g,h}$ become constant gauge transformations
defining elements of $H^2(\Gamma, U(1))$.

\subsection{Orbifold group actions on $B$ fields}

In this section we shall work out a description of elements of
the set of orbifold group actions on $B$ fields.
As for $U(1)$ gauge fields, orbifold group actions on $B$ fields
can be determined by specifying how pullbacks are related to original
data, so we shall be studying pullbacks.

It is important to emphasize at the start, that we really do mean
to use the word set -- the set of orbifold group actions can, in
general, not canonically be given any group structure.
Most mathematically oriented physics papers calculate
cohomology groups or generalized cohomology groups -- here, by
contrast, we shall begin by calculating a set, which cannot
be understood as a group in general, much less any sort of
(generalized) cohomology group.
After we have worked out this set, in the next section we shall
argue that elements of this set differ by
gauge transformations of $B$ fields, which will lead us
to discover $H^2(\Gamma, U(1))$.

Let $\{ U_{\alpha} \}$ be a ``good invariant'' cover, as before.
Then, a two-form field potential is described as a collection 
of two-forms $B^{\alpha}$, one for each open set $U_{\alpha}$,
related by gauge transformation on overlaps.

We assume that the $B$ field has no magnetic sources
(meaning, that the exterior derivative of its curvature $H$ vanishes),
and that the curvature $H$ is (the image of) an element of
integral cohomology $H^3({\bf Z})$.
Such $B$ fields on a smooth space $X$ are described on the
open cover $\{ U_{\alpha} \}$ by \cite{hitchin,cjthesis}
two-forms $B^{\alpha}$ on $U_{\alpha}$,
one-forms $A^{\alpha \beta}$ on $U_{\alpha} \cap U_{\beta} =
U_{\alpha \beta}$, and $U(1)$-valued functions $h_{\alpha
\beta \gamma}$ on $U_{\alpha \cap \beta \cap \gamma}
= U_{\alpha \beta \gamma}$, satisfying
\begin{eqnarray*}
B^{\alpha} \: - \: B^{\beta} & = & d A^{\alpha \beta} \\
A^{\alpha \beta} \: + \: A^{\beta \gamma} \: + \:
A^{\gamma \alpha} & = & d \, \log h_{\alpha \beta \gamma} \\
\delta ( h_{\alpha \beta \gamma} ) & = & 1
\end{eqnarray*}
We should mention that in writing the above we are not putting 
any more structure
on $B$ fields than is already present in string theory.
For example, the reader might be concerned at the appearance of vector fields
$A^{\alpha \beta}$; such a reader should be reminded that if we
define $B$ fields in open patches, then on overlaps the $B$ fields
will differ by some gauge transformation.  By specifying the 
$A^{\alpha \beta}$ we have merely made the gauge
transformations on overlaps explicit, no more.

In this section we shall work out how to describe orbifold group
actions on $B$-fields, in terms of the data above.
To begin, demand that the \v{C}ech cocycles $h_{\alpha \beta \gamma}$
are preserved by the orbifold group, up to coboundaries.
Specifically, demand
\begin{equation}     \label{nudefn}
g^* h_{\alpha \beta \gamma} \: = \: h_{\alpha \beta \gamma} \,
\nu^g_{\alpha \beta} \, 
\nu^g_{\beta \gamma} \,
\nu^g_{\gamma \alpha}
\end{equation}
for some \v{C}ech cochains $\nu^g$, for each $g \in \Gamma$.

Next, we shall derive a constraint on the coboundaries
$\nu^g$ from self-consistency of equation~(\ref{nudefn}).
Specifically, 
\begin{eqnarray*}
(g_1 g_2)^* h_{\alpha \beta \gamma} & = &
 h_{\alpha \beta \gamma} \,
\nu^{g_1 g_2}_{\alpha \beta} \, 
\nu^{g_1 g_2}_{\beta \gamma} \,
\nu^{g_1 g_2}_{\gamma \alpha} \\
\mbox{ also } & = & g_2^* \left( \,
h_{\alpha \beta \gamma} \,
\nu^{g_1}_{\alpha \beta} \, 
\nu^{g_1}_{\beta \gamma} \,
\nu^{g_1}_{\gamma \alpha} \, \right) \\
& = & h_{\alpha \beta \gamma} \, 
\left( \nu^{g_2}_{\alpha \beta} \, g_2^* \nu^{g_1}_{\alpha \beta} \right) \,
\left( \nu^{g_2}_{\beta \gamma} \, g_2^* \nu^{g_1}_{\beta \gamma} \right) \,
\left( \nu^{g_2}_{\gamma \alpha} \, g_2^* \nu^{g_1}_{\gamma \alpha} \right)
\end{eqnarray*}
from which we see that
\begin{equation}    \label{hdefn}
\nu^{g_1 g_2}_{\alpha \beta} \: = \:
\nu^{g_2}_{\alpha \beta} \, g_2^* \nu^{g_1}_{\alpha \beta} \,
\left( h^{g_1, g_2}_{\alpha} \right) \,
\left( h^{g_1, g_2}_{\beta} \right)^{-1}
\end{equation}
for some \v{C}ech cochains $h^{g_1, g_2}$.

By applying equation~(\ref{hdefn}) to expand $\nu^{g_1 g_2 g_3}$
in two different ways, we can derive
\begin{equation}   \label{hreln}
\left( h^{g_1, g_2 g_3}_{\alpha} \right)
\, \left( h^{g_2, g_3}_{\alpha} \right) \: = \:
\left( g_3^* h^{g_1, g_2}_{\alpha} \right) \, 
\left( h^{g_1 g_2, g_3}_{\alpha} \right)
\end{equation}

Next, consider the forms $B^{\alpha}$ and $A^{\alpha \beta}$.
Define two-forms $B(g)^{\alpha}$ and one-forms $A(g)^{\alpha}$
by
\begin{eqnarray*}
g^* B^{\alpha} & = & B^{\alpha} \: + \: B(g)^{\alpha} \\
g^* A^{\alpha \beta} & = & A^{\alpha \beta} \: + \:
A(g)^{\alpha \beta}
\end{eqnarray*}
We shall use self-consistency to work out meaningful expressions for
$B(g)^{\alpha}$ and $A(g)^{\alpha \beta}$.

From the expression
\begin{displaymath}
g^*\left( \, B^{\alpha} \, - \, B^{\beta} \, \right) \: = \:
g^* d A^{\alpha \beta}
\end{displaymath}
one can derive
\begin{equation}    \label{b1}
B(g)^{\alpha} \: - \: B(g)^{\beta} \: = \: d A(g)^{\alpha \beta}
\end{equation}
Furthermore, by expanding both sides of the expression
\begin{displaymath}
g^* \left( \, A^{\alpha \beta} \: + \: A^{\beta \gamma} \: + \:
A^{\gamma \alpha} \, \right) \: = \:
g^* d \, \log h_{\alpha \beta \gamma}
\end{displaymath}
we find
\begin{equation}    \label{b2}
A(g)^{\alpha \beta} \: = \: d \, \log \nu^g_{\alpha \beta} \: + \:
\Lambda(g)^{\alpha} \: - \: \Lambda(g)^{\beta}
\end{equation}
for some one-forms $\Lambda(g)^{\alpha}$ defined on open sets $U_{\alpha}$.
Comparing equations~(\ref{b1}) and (\ref{b2}), we find
\begin{equation}
B(g)^{\alpha} \: = \: d \Lambda(g)^{\alpha}
\end{equation}

By expanding $(g_1 g_2)^* B^{\alpha}$ in two different
ways, we find
\begin{equation}
\Lambda(g_1 g_2)^{\alpha} \: = \:
\Lambda(g_2)^{\alpha} \: + \: g_2^* \Lambda(g_1)^{\alpha}
\: + \: d \Lambda^{(2)}(g_1, g_2)^{\alpha}
\end{equation}
for some real-valued functions $\Lambda^{(2)}(g_1, g_2)^{\alpha}$
defined on the open sets $U_{\alpha}$.

Finally, by expanding $(g_1 g_2)^* A^{\alpha \beta}$ in two different
ways, we find that
\begin{displaymath}
d \Lambda^{(2)}(g_1, g_2)^{\alpha} \: = \: - d \log h^{g_1, g_2}_{\alpha}
\end{displaymath}

We can summarize the results of these computations as follows:
\begin{eqnarray*}
g^* B^{\alpha} & = & B^{\alpha} \: + \: d \Lambda(g)^{\alpha} \\
g^* A^{\alpha \beta} & = & A^{\alpha \beta} \: + \:
d \log \nu^g_{\alpha \beta} \: + \:
\Lambda(g)^{\alpha} \: - \: \Lambda(g)^{\beta} \\
\Lambda(g_1 g_2)^{\alpha} & = & \Lambda(g_2)^{\alpha} \: + \:
g_2^* \Lambda(g_1)^{\alpha} \: - \: d \log h^{g_1, g_2}_{\alpha} \\
g^* h_{\alpha \beta \gamma} & = & h_{\alpha \beta \gamma} \,
\nu^g_{\alpha \beta} \, \nu^g_{\beta \gamma} \, \nu^g_{\gamma \alpha} \\
\nu^{g_1 g_2}_{\alpha \beta} & = & \left( \nu^{g_2}_{\alpha \beta}
\right) \, \left( g_2^* \nu^{g_1}_{\alpha \beta} \right) \,
\left( h^{g_1, g_2}_{\alpha} \right) \,
\left( h^{g_1. g_2}_{\beta} \right)^{-1} \\
\left( h^{g_1, g_2 g_3}_{\alpha} \right) \, 
\left( h^{g_2, g_3}_{\alpha} \right) & = &
\left( g_3^* h^{g_1, g_2}_{\alpha} \right) \, 
\left( h^{g_1 g_2, g_3}_{\alpha} \right)
\end{eqnarray*}
where $\Lambda(g)^{\alpha}$, $\nu^g_{\alpha \beta}$,
and $h^{g_1, g_2}_{\alpha}$ are structures introduced to define
the action of the orbifold group on the $B$ field.

We have defined orbifold group actions on $B$ fields
in terms of the transition functions and other local data
defining the $B$ field.  More formally, a $B$ field
can also be understood as a connection on a ``1-gerbe,''
a special kind of stack, or (loosely) sheaf of categories.
We discussed such objects in \cite{dt2}, together with 
a discussion of how one defines orbifold group actions on them.
Our discussion in \cite{dt2} is adapted from \cite{brylinski},
which discusses $B$ fields in the language of stacks.

\subsection{Differences between orbifold group actions}

In describing orbifold $U(1)$ Wilson lines, the group
$H^1(\Gamma, U(1))$ arises as differences between orbifold group actions;
similarly, in describing discrete torsion, $H^2(\Gamma, U(1))$ arises
in describing the differences between orbifold group actions.
In both cases, one can get any action from any other action 
by combining the action with a set of gauge transformations;
in the former case, $H^1(\Gamma, U(1))$ counts those gauge transformations,
and in the latter case, $H^2(\Gamma, U(1))$ counts some of the gauge
transformations.

One unusual matter we shall discover in this section is that,
in addition to elements of $H^2(\Gamma, U(1))$, one sometimes
has additional actions of the orbifold group on the $B$ fields
 -- sometimes, there may be more to ``discrete torsion'' than just
$H^2(\Gamma, U(1))$.  We shall show in a later section that, at the
level of twisted sector contributions to partition functions,
these additional contributions can do more than merely multiply
contributions $Z_{(g,h)}$ by a phase, they significantly alter
the $Z_{(g,h)}$ themselves.  We shall discuss these elements
in much greater detail later in this section.

Consider two distinct orbifold group actions on the $B$ fields.
Denote one orbifold group action as in the last section,
and denote the second action with a bar, e.g., $\overline{\nu}^g_{\alpha
\beta}$ rather than $\nu^g_{\alpha \beta}$.

Define
\begin{equation}
T^g_{\alpha \beta} \: = \: \frac{  \nu^g_{\alpha \beta} }{
\overline{\nu}^g_{\alpha \beta} }
\end{equation}
From dividing the expressions
\begin{eqnarray*}
g^* h_{\alpha \beta \gamma} & = & \left( h_{\alpha \beta \gamma} \right) \,
             \left( \nu^g_{\alpha \beta} \right) \,
             \left( \nu^g_{\beta \gamma} \right) \,
             \left( \nu^g_{\gamma \alpha} \right) \\
\mbox{ also } & = & \left( h_{\alpha \beta \gamma} \right) \,
             \left( \overline{\nu}^g_{\alpha \beta} \right) \,
             \left( \overline{\nu}^g_{\beta \gamma} \right) \,
             \left( \overline{\nu}^g_{\gamma \alpha} \right) 
\end{eqnarray*}
we see that
\begin{displaymath}
T^g_{\alpha \beta} \, T^g_{\beta \gamma} \,
T^g_{\gamma \alpha} \: = \: 1
\end{displaymath}
meaning that the $T^g$ are transition functions for a principal
$U(1)$ bundle.

The fact that we are seeing principal $U(1)$ bundles at the same
place where gauge transformations appeared in describing
orbifold $U(1)$ Wilson lines is no accident -- a gauge
transformation of a set of $B$ fields is defined by
a(n equivalence class of) principal $U(1)$ bundles with connection
\cite{dt2}.  So, just as for orbifold Wilson lines,
we are already seeing that the difference between two lifts
is defined by a gauge transformation -- the only difference being
that for $B$ fields, a ``gauge transformation'' is defined by a bundle.

Next, define
\begin{equation}
\omega^{g,h}_{\alpha} \: = \: \frac{ h^{g,h}_{\alpha} }{
\overline{h}^{g,h}_{\alpha} }
\end{equation}
From dividing the expressions
\begin{eqnarray*}
\nu^{gh}_{\alpha \beta} & = & 
\left( \nu^h_{\alpha \beta} \right) \,
\left( h^* \nu^g_{\alpha \beta} \right) \,
\left( h^{g,h}_{\alpha} \right) \,
\left( h^{g,h}_{\beta} \right)^{-1} \\
\overline{\nu}^{gh}_{\alpha \beta} & = &
\left( \overline{\nu}^h_{\alpha \beta} \right) \,
\left( h^* \overline{\nu}^g_{\alpha \beta} \right) \,
\left( \overline{h}^{g,h}_{\alpha} \right) \,
\left( \overline{h}^{g,h}_{\beta} \right)^{-1}
\end{eqnarray*}
we find that
\begin{equation}
T^{gh}_{\alpha \beta} \: = \:
\left( T^{h}_{\alpha \beta} \right) \,
\left( h^* T^g_{\alpha \beta} \right) \,
\left( \omega^{g,h}_{\alpha} \right) \,
\left( \omega^{g,h}_{\beta} \right)^{-1}
\end{equation}
which means that the $\omega^{g,h}_{\alpha}$ are
local-coordinate realizations \cite[section 5.5]{husemoller}
of a map $\omega^{g,h}$ between bundles:
\begin{equation}
\omega^{g,h}: \: T^h \otimes h^* T^g \: \longrightarrow \: T^{gh}
\end{equation}

We should take a moment to carefully explain what we mean by
$\otimes$, since we have been describing the bundles $T^g$ as principal
bundles, not vector bundles.  One way to understand $\otimes$ is
to think of it as the product of abelian torsors, following
\cite[section 5.1]{brylinski}.  Alternatively, one could think
of the $T^g$ as complex line bundles with hermitian fiber metrics.
Perhaps the easiest way to understand $\otimes$ in the present
context is simply as, the bundle whose transition functions are
the product of the transition functions of the bundles appearing
in the product.

Next, from 
\begin{displaymath}
\left( h^{g_1, g_2 g_3}_{\alpha} \right) \,
\left( h^{g_2, g_3}_{\alpha} \right) 
\: = \:
\left( g_3^* h^{g_1, g_2}_{\alpha} \right) \,
\left( h^{g_1 g_2, g_3}_{\alpha} \right)
\end{displaymath}
we derive the commutivity condition
\begin{displaymath}
\begin{array}{ccc}
T^{g_3} \otimes g_3^* \left( \, T^{g_2} \otimes g_2^* T^{g_1} \right)
& \stackrel{ \omega^{g_1, g_2} }{ \longrightarrow } &
T^{g_3} \otimes g_3^* T^{g_1 g_2} \\
\makebox[0pt][r]{ $\scriptstyle{  \omega^{g_2, g_3} }$} \downarrow
& & \downarrow \makebox[0pt][l]{
$\scriptstyle{ \omega^{g_1 g_2, g_3} }$ } \\
T^{g_2 g_3} \otimes (g_2 g_3)^* T^{g_1} &
\stackrel{ \omega^{g_1, g_2 g_3} }{\longrightarrow } &
T^{g_1 g_2 g_3}
\end{array}
\end{displaymath}

To summarize our results so far, we have found that
any two orbifold group actions on the same set of $B$ fields
differ by a set of principal $U(1)$ bundles $T^g$ together
with bundle morphisms $\omega^{g_1, g_2}: T^{g_2} \otimes g_2^* T^{g_1}
\rightarrow T^{g_1 g_2}$.  As mentioned before, this is to say that
any two orbifold group actions on $B$ fields differ by a gauge
transformation, as a principal $U(1)$ bundle defines a gauge
transformation of $B$ fields.

Next, define 
\begin{equation}
\Xi(g)^{\alpha} \: = \:
\overline{\Lambda}(g)^{\alpha} \: - \:
\Lambda(g)^{\alpha}
\end{equation}
From subtracting the expressions
\begin{eqnarray*}
g^* A^{\alpha \beta} & = & A^{\alpha \beta} \: + \:
d \log \nu^g_{\alpha \beta} \: + \:
\Lambda(g)^{\alpha} \: - \: \Lambda(g)^{\beta} \\
\mbox{ also } & = & A^{\alpha \beta} \: + \:
d \log \overline{\nu}^g_{\alpha \beta} \: + \:
\overline{\Lambda}(g)^{\alpha} \: - \:
\overline{\Lambda}(g)^{\beta}
\end{eqnarray*}
we find that
\begin{equation}
\Xi(g)^{\alpha} \: - \: \Xi(g)^{\beta} \: = \:
d \log T^g_{\alpha \beta}
\end{equation}
In other words, the local one-forms $\Xi(g)^{\alpha}$ define a connection
on the bundle $T^g$.

From subtracting
\begin{eqnarray*}
g^* B^{\alpha} & = & B^{\alpha} \: + \: d \Lambda(g)^{\alpha} \\
\mbox{ also } & = & B^{\alpha} \: + \: d \overline{\Lambda}(g)^{\alpha}
\end{eqnarray*}
we see that
\begin{equation}
d \Xi(g)^{\alpha} \: = \: 0
\end{equation}
In other words, the connection $\Xi(g)^{\alpha}$ is not just any
connection on the principal $U(1)$ bundle $T^g$, but must be a flat
connection.  The analogous statement in studying orbifold group
actions on $U(1)$ gauge fields is that any two orbifold group
actions differ by a constant gauge transformation.

From subtracting
\begin{eqnarray*}
\Lambda(g_1 g_2)^{\alpha} & = & \Lambda(g_2)^{\alpha} \: + \:
g_2^* \Lambda(g_1)^{\alpha} \: - \: d \log h^{g_1, g_2}_{\alpha} \\
\overline{\Lambda}(g_1 g_2)^{\alpha} & = &
\overline{\Lambda}(g_2)^{\alpha} \: + \:
g_2^* \overline{\Lambda}(g_1)^{\alpha} \: - \:
d \log \overline{h}^{g_1, g_2}_{\alpha}
\end{eqnarray*}
we find that
\begin{equation}   \label{xiomega}
\Xi(g_1 g_2)^{\alpha} \: = \:
\Xi(g_2)^{\alpha} \: + \: g_2^* \Xi(g_1)^{\alpha} \: + \:
d \log \omega^{g_1, g_2}_{\alpha}
\end{equation}
which means that the bundle morphism $\omega^{g_1, g_2}:
T^{g_2} \otimes g_2^* T^{g_1} \rightarrow T^{g_1 g_2}$
is constrained to preserve the connection on the bundles.

To summarize our progress so far, we have found that any two
orbifold group actions on a set of $B$ fields differ by a collection
of principal $U(1)$ bundles $T^g$ with flat connection $\Xi(g)$,
together with connection-preserving bundle morphisms
$\omega^{g_1, g_2}: T^{g_2} \otimes g_2^* T^{g_1} \rightarrow
T^{g_1 g_2}$, such that the following diagram commutes:
\begin{equation}    \label{omegacocycle}
\begin{array}{ccc}
T^{g_3} \otimes g_3^* \left( \, T^{g_2} \otimes g_2^* T^{g_1} \right)
& \stackrel{ \omega^{g_1, g_2} }{ \longrightarrow } &
T^{g_3} \otimes g_3^* T^{g_1 g_2} \\
\makebox[0pt][r]{ $\scriptstyle{  \omega^{g_2, g_3} }$} \downarrow
& & \downarrow \makebox[0pt][l]{
$\scriptstyle{ \omega^{g_1 g_2, g_3} }$ } \\
T^{g_2 g_3} \otimes (g_2 g_3)^* T^{g_1} &
\stackrel{ \omega^{g_1, g_2 g_3} }{\longrightarrow } &
T^{g_1 g_2 g_3}
\end{array}
\end{equation}

We should be careful at this point.  Although we have
not emphasized this point, it is only equivalence
classes of bundles $T^g$ with connection $\Xi(g)$ that are relevant.
In a nutshell, if $\Lambda$ and $\Lambda'$ are two one-forms that
differ by an exact form, then $B + d \Lambda = B + d \Lambda'$,
so any two bundles with connection that differ by a gauge transformation
(of the bundle) define the same action on the $B$ field.
So, if $T'^g$ is another set of principal $U(1)$ bundles
with connection $\Xi'(g)$, and $\kappa_g: T^g \rightarrow T'^g$
are connection-preserving bundle isomorphisms,
then we can replace the data given above with the collection
$(T'^g, \Xi'(g), \omega'^{g_1, g_2})$, where the $\omega'$ are
given by
\begin{equation}    \label{omegacobound}
\omega'^{g_1, g_2} \: \equiv \:
\kappa_{g_1 g_2} \circ \omega^{g_1, g_2} \circ
\left( \kappa_{g_2} \otimes g_2^* \kappa_{g_1} \right)^{-1}
\end{equation}
to get an equivalent orbifold group action on the $B$ fields.

\subsection{$H^2(\Gamma, U(1))$}

How do elements of $H^2(\Gamma, U(1))$ arise?
Take the bundles $T^g$ to be topologically trivial,
and the connections $\Xi(g)$ to be gauge-trivial.
We can then map the bundles $T^g$ to the canonical trivial
bundle (whose transition functions are all identically $1$),
and gauge-transform the connections $\Xi(g)$ to zero.

In this case, the bundle morphisms $\omega^{g_1, g_2}:
T^{g_2} \otimes g_2^* T^{g_1}$ become gauge transformations
of the canonical trivial bundle.  From the fact that
the $\omega^{g_1, g_2}$ must preserve the connection
(i.e., equation~(\ref{xiomega})), and assuming the covering
space is connected, we see that the gauge transformations
$\omega^{g_1, g_2}$ must be constant gauge transformations.

We have reduced the data describing this set of $B$ field
gauge transformations to a set of maps $\omega: \Gamma \times \Gamma
\rightarrow U(1)$.  From commutivity of diagram~(\ref{omegacocycle}),
we see that the $\omega$ define a group 2-cocycle, i.e.,
\begin{displaymath}
\omega^{g_1, g_2 g_3} \, \omega^{g_2, g_3} \: = \:
\omega^{g_1 g_2, g_3} \, \omega^{g_1, g_2}
\end{displaymath}

So far we have reduced the data describing this set of $B$ field
gauge transformations to a group 2-cocycle.  More can be said:
there is still a residual set of gauge transformations that 
must be taken into account.  We can perform a constant gauge
transformation on each of the bundles $T^g$; this will preserve
the connection on each bundle.  From equation~(\ref{omegacobound}),
we see that these constant gauge transformations simply change the
group 2-cocycle $\omega^{g_1, g_2}$ by a coboundary.

Thus, we find that this set of $B$ field gauge transformations
is classified by elements of $H^2(\Gamma, U(1))$.

It should be emphasized that the appearance of $H^2(\Gamma, U(1))$
above holds regardless of whether or not $\Gamma$ acts freely -- nowhere
have we made any assumptions concerning how the orbifold group $\Gamma$
acts.

\subsection{Detailed classification of orbifold group actions}

In the previous subsection we showed how elements
of $H^2(\Gamma, U(1))$ describe at least some differences
between orbifold group actions.  For $U(1)$ gauge fields,
we found that all orbifold group actions differed by some
element of $H^1(\Gamma, U(1))$ -- to what extent can the analogous
statement be made here?

Do all orbifold group actions on $B$ fields differ by
such data described by $H^2(\Gamma, U(1))$? 
We shall argue that under special circumstances, 
all orbifold group actions on $B$ fields differ by
elements of $H^2(\Gamma, U(1))$, but in general there can
be additional differences.

Suppose the covering space $X$ is connected, simply-connected,
and $H^2(X,{\bf Z})$ has no torsion.
Let $(T^g, \Xi(g), \omega^{g_1, g_2})$ be a set of data defining
the difference between two orbifold group actions.
We know that the connections $\Xi(g)$ on the bundles $T^g$ are
flat, which means that for each $g$, $c_1(T^g)$ must be a torsion
element of $H^2(X, {\bf Z})$.  However, we have assumed that $H^2(X,{\bf Z})$
has no torsion -- so the bundles $T^g$ must all be topologically
trivial, i.e., $c_1(T^g) = 0$ in $H^2(X, {\bf Z})$.
In addition, we assumed the space $X$ is simply-connected.
On a simply-connected space, the only flat connections on a topologically
trivial bundle are gauge-trivial (gauge-equivalent to the zero connection).
So, if $X$ is simply-connected and $H^2(X, {\bf Z})$ has no torsion,
then the bundles $T^g$ are all topologically trivial and the
connections $\Xi(g)$ are all gauge-trivial.  As noted in the
last section, such gauge transformations of $B$ fields
are classified by $H^2(X, {\bf Z})$.

Thus, if the covering space $X$ is simply-connected and $H^2(X, {\bf Z})$
has no torsion, then any two orbifold group actions on a $B$ field
differ by a set of gauge transformations classified by an element
of $H^2(\Gamma, U(1))$.

Suppose now that these criteria are not met -- $X$ is not simply-connected,
or $H^2(X, {\bf Z})$ has torsion.  Then not all $B$-field-gauge-transformations
need be described by topologically trivial bundles $T^g$ with
gauge trivial connections $\Xi(g)$ -- if $X$ is not simply-connected,
then even on a topologically trivial bundle one can have flat connections
which are not gauge-trivial, and if $H^2(X, {\bf Z})$ has torsion,
then one can have topologically nontrivial bundles with flat connections.

As a result, in general it appears that there can be additional orbifold
group actions on $B$ fields, beyond those classified by $H^2(\Gamma, U(1))$.

In retrospect, we should not have been surprised.  Consider the
special case in which $\Gamma$ is freely-acting and the $B$ field
is identically zero everywhere.  We can compute the homology
of the quotient $X/\Gamma$ from the Cartan-Leray spectral sequence
\cite[section VII.7]{brown}:
\begin{displaymath}
E^2_{p,q} \: = \: H_p \left( \Gamma, H_q(X, {\bf Z}) \right)
\: \Longrightarrow \: H_{p+q}(X/\Gamma, {\bf Z})
\end{displaymath}
so (loosely ignoring differentials for simplicity) we see that
$H_2(X/\Gamma, {\bf Z})$ ultimately receives contributions from not only
$H_2(\Gamma, {\bf Z})$ (which dualizes to $H^2(\Gamma, U(1))$)
and $H_2(X, {\bf Z})$, but also from $H_1 \left(\Gamma, H_1(X, 
{\bf Z}) \right)$
 -- so if $X$ is not simply connected, then one should expect additional
contributions beyond those determined by $H^2(\Gamma, U(1))$
and $H_2(X, {\bf Z})$.
(For a lengthier discussion of such 
cohomology calculations, and how $H^2(\Gamma, U(1))$ enters them,
see \cite{dt1}.)

How should such additional orbifold group actions show
up physically?
In terms of twisted sector contributions to one-loop partition functions,
for example, the orbifold group actions classified by
$H^2(\Gamma, U(1))$ merely multiply twisted-sector contributions
$Z_{(g,h)}$ by phases.  We shall argue later that these additional
orbifold group actions do more than just multiply $Z_{(g,h)}$ by a phase
 -- these will deeply change $Z_{(g,h)}$ itself, by changing the
weighting of individual sigma model map contributions to $Z_{(g,h)}$
in a winding-number-dependent fashion.

In fact, we shall argue in much greater detail 
elsewhere \cite{dtshift} that these `new'
degrees of freedom are actually some very old degrees of freedom,
the so-called `shift orbifolds' that play an important role
in asymmetric orbifolds, but are rather more boring in symmetric
orbifolds.

\subsection{Commentary}

So far we have described the set of orbifold group actions
on $B$ fields, described differences between any two orbifold
group actions on a fixed $B$ field, unveiled $H^2(\Gamma, U(1))$,
and also discovered some new and more subtle orbifold group actions.

A few general comments are in order.

First, the results of this section do not depend upon $\Gamma$
being freely acting.  Everything we have described is the same
for $\Gamma$ having fixed points as for $\Gamma$ freely acting
-- the details of the action of $\Gamma$ on the base space
are entirely irrelevant.

Second, the results of this section do not depend upon whether
or not $\Gamma$ is abelian.  We get the same results if $\Gamma$
is nonabelian.

Third, the results of this section do not depend upon whether
or not the $B$ field is flat or topologically trivial.
In principle, precisely the same remarks hold regardless of whether
$H = 0$ or $H \neq 0$ in $H^3({\bf Z})$ on the covering space.
However, if the $B$ fields are described by a nonzero element of
$H^3({\bf Z})$ -- if the 1-gerbe is not topologically trivial -- then
one must check whether an orbifold group action actually exists, just as
for orbifold $U(1)$ Wilson lines.  

Now, we have described $H^2(\Gamma, U(1))$ as arising in the differences
between two orbifold group actions, but that is not quite how
people usually discuss it -- people speak of ``turning on'' 
discrete torsion.  This is because, just as for orbifold $U(1)$ Wilson
lines, in almost every case in physics where orbifolds are studied,
the $B$ fields are such that there is a canonical trivial orbifold
group action.  For example, just as for orbifold $U(1)$ Wilson lines,
if the $B$ field is topologically trivial, then there is a notion
of a canonical trivial action, and so we can define any orbifold group
action in terms of its difference from the trivial action.
Thus, in this case, we can indeed ``turn on'' discrete torsion.

In general, however, one can not always expect to have such
a canonical trivial lift.  
This matter is discussed further in \cite{dt1}.

\section{Derivation of twisted sector phases}   \label{twphase}

In this section we shall derive the phases that appear
in twisted sector contributions to partition functions,
as originally described in \cite{vafa1}.

These phases appear precisely because the string sigma model
has a term
\begin{displaymath}
\int \, B
\end{displaymath}
On the covering space, a contribution to a twisted sector
is a polygon with sides identified under the group action.
The group action that identifies the sides lifts to
an action on the B field; that action contributes a phase
to
\begin{displaymath}
\exp \left( \, \int \, B \, \right)
\end{displaymath}
and so the twisted sector contribution to the partition function
comes with a phase.

We shall begin in section~\ref{owlanalog} by discussing 
a simpler analogue of this behavior for orbifold Wilson lines.
In section~\ref{1loop} we shall derive Vafa's twisted sector phases
for string one-loop partition functions, and in section~\ref{2loop}
we shall derive the twisted sector phases for string two-loop 
partition functions.

\subsection{Analogue for orbifold $U(1)$ Wilson lines}   \label{owlanalog}

\begin{figure}
\centerline{\psfig{file=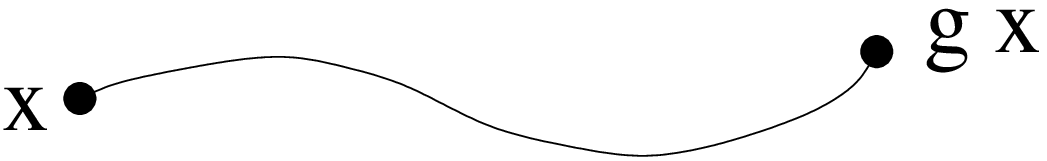,width=2in}}
\caption{  \label{figWline}
Lift of closed loop to covering space.}
\end{figure}

To root ourselves, we shall begin by reviewing the analogue
of twisted sector phases for orbifold Wilson lines.
Consider computing a Wilson loop in some gauge theory
(for simplicity, we shall assume a $U(1)$ gauge theory) on the quotient
space.  Suppose furthermore that such a loop descends from an open
loop on the covering space, whose ends are identified by the
action of some element $g$ of the orbifold group $\Gamma$, as
shown in figure~(\ref{figWline}).  For simplicity we shall
assume the bundle on which the connection lives is topologically
trivial, and so has a canonical trivial orbifold group action --
so we can specify any other orbifold group action in terms of
a set of gauge transformations.

How do we calculate the value of the Wilson loop
\begin{displaymath}
\exp \left( \, \int \, A \, \right)
\end{displaymath}
while working on the covering space?

In order to see what to do, first consider the problem of calculating
the holonomy of a vector connection, in the case that
$A^{\alpha}$ is only defined in patches along the loop.
In such a case, one splits the loop into segments,
each segment completely contained in a patch in which $A^{\alpha}$
is defined.  Then, the holonomy is given by the product
of $\exp( \int A )$ from each patch, separated by factors of the
transition functions (evaluated at the borders of the
segments).  At the end of the day, one can check
that this holonomy is independent of the precise splitting of the
loop into segments.

Now, we shall return to the problem of calculating Wilson loops
on covering spaces.
First, there is a contribution to the Wilson loop from
integrating the vector field $A$ along the path from
$x$ to $g x$, i.e., there is a factor
\begin{equation}      \label{naiveWline}
\exp \left( \, \int^{gx}_x \, A \, \right)
\end{equation}

However, this factor is not the end of the story.
A Wilson loop (i.e., a Wilson line around a closed loop) on the quotient
space will be invariant under gauge transformations,
whereas the factor~(\ref{naiveWline}) does not appear invariant at all.
To fix matters, use the relationship between the gauge field $A$
at $x$ and at $g x$, i.e.,
\begin{displaymath}
A_{gx} \: = \: g^* A_x \: = \: A_x \: + \: d \log \varphi^g
\end{displaymath}
where $\varphi^g$ is a $U(1)$-valued function defining
a gauge transformation -- rather, defining the action of the
orbifold group.

To close the loop, we need to include the gauge transformation relating
$A_{gx}$ and $A_x$.  The correct value of the Wilson
loop, as calculated on the covering space, is
\begin{equation}  \label{trueWline}
\varphi^g_x \, \exp \left( \, \int^{gx}_x \, A \, \right)
\end{equation}

We shall follow a similar procedure in analyzing the phase factor
$\exp \left( \int B \right)$ in twisted sectors.
The naive integral of $B$ over a polygon in the covering space
is not sufficient; we must add gauge transformations along the
boundary, which in this case are Wilson lines along the boundary.
Furthermore, for $B$ fields, those Wilson lines along the boundary
are still not quite sufficient -- we also need to account for
bad behavior at the corners of the polygon.

\subsection{One loop}   \label{1loop}

\begin{figure}   
\centerline{\psfig{file=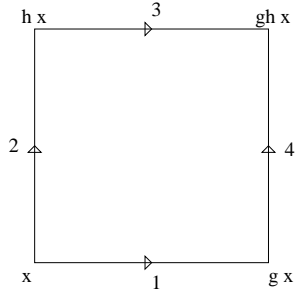,width=1.5in}}
\caption{  \label{fig1loop}
A twisted sector
contribution to the one-loop partition function.}
\end{figure}

The string orbifold one-loop partition function receives contributions
not only from $T^2$'s in the covering space, but also from
configurations of strings that form $T^2$'s on the quotient space,
but only form open polygons on the covering space, as illustrated
in figure~(\ref{fig1loop}).

Just as orbifold Wilson loops received an extra phase when
edges are identified by group actions, so ``orbifold Wilson surfaces''
\begin{displaymath}
\exp \, \left( \, \int \, B \, \right)
\end{displaymath}
get boundary contributions due to the orbifold action.
In other words, naively integrating $B$ over the region indicated
will not give the phase over the cycle on the quotient space -- we
must also include contributions induced by gauge transformations of
the $B$ field occurring at the boundaries as the edges are glued 
together.

Before we begin the analysis, let us briefly pause to
review how one calculates the holonomy of a $B$ field over
a surface, in the event that there is no two-form defined everywhere
over the surface, but instead $B$ is only defined on patches.
To calculate the holonomy in this case, first tile the surface,
in such a way that each tile is contained within a patch
on which $B^{\alpha}$ is known, and each component of each tile boundary
is contained within an overlap of patches.
Then the holonomy of the $B$ field is a product of several factors:
\begin{enumerate}
\item For each tile, there is a factor of $\exp( \int B^{\alpha} )$,
computed using the $B$ field associated to the patch in which the
tile lies.
\item For each component of each tile boundary, there is a factor of
$\exp( \int A^{\alpha \beta} )$, where $B^{\alpha} - B^{\beta} = d 
A^{\alpha \beta}$.
\item Finally, one can check that in order for the resulting holonomy
to be independent of the choice of tiling, one must add factors
of\footnote{Where $\delta A^{\alpha \beta} = d \log h_{\alpha \beta
\gamma}$.} $h_{\alpha \beta \gamma}$, evaluated at each point where multiple
boundary components intersect.
\end{enumerate}
At the end of the day, the result is independent of the
choice of tiling.  
(Note that, when applied to WZW models, this gives us a means of 
understanding the exponential of the Wess-Zumino term that does not
involve appealing to bounding three-manifolds -- after all,
the exponential of the Wess-Zumino term is just the holonomy of 
the pullback of the $B$ field on the group manifold.  The fact that
the Wess-Zumino term can change by integral amounts comes from
possible gauge transformations of the $B$ field.)
After reflection, it is clear that a closely related
procedure will allow us to calculate the orbifold Wilson surface
holonomies, as we shall now demonstrate.

\subsubsection{Basic analysis}

To be specific, consider the region $D$ shown in figure~(\ref{fig1loop}).
Sides 1 and 3 are identified by $h \in \Gamma$,
and sides 2 and 4 are identified by $g \in \Gamma$.
As a technical aside, note that, regardless of 
whether $\Gamma$ is abelian, one sums
over contributions from commuting pairs $(g,h)$ in describing
twisted sectors, precisely so that the square shown in
figure~(\ref{fig1loop}) will close at the upper right.

For simplicity we shall assume that\footnote{Also assume
the associated gerbe is topologically trivial.} $B \equiv 0$.
In this case, since there is a canonical trivial action of the 
orbifold group on the $B$ fields, we can define any other action
of the orbifold group on the $B$ fields entirely in terms of
the gauge transformations that distinguish it from the canonical
trivial action.
Let such a set of gauge transformations be denoted by
bundles $T^g$ with connection $\Lambda(g)$, and connecting
bundle maps $\omega^{g_1, g_2}$.

For the moment, we shall assume that the bundles $T^g$ are the canonical 
trivial bundle (so the $\omega^{g_1, g_2}$ are all merely gauge
transformations, not bundle morphisms), 
and the connections $\Lambda(g)$ are all gauge-equivalent
to the zero connection.  We shall examine the more general case
after examining this case in detail.

Now,
under a gauge transformation by a bundle $T$ with connection\footnote{
Previously we denoted the connection on bundle $T^g$ by $\Xi(g)$,
and used $\Lambda(g)$ to denote data used to define the action
of the orbifold group on a $B$ field.  At this point we
are changing notation -- we will use $\Lambda(g)$ instead of
$\Xi(g)$ to denote the
connection on bundle $T^g$.}
$\Lambda$, the $B$ field locally transforms as
\begin{displaymath}
B \: \mapsto \: B \: + \: d \Lambda
\end{displaymath}
and so the holonomy of $B$ over some surface $S$ transforms as
\begin{displaymath}
\exp \, \left( \, \int_S \, B \, \right)
\: \mapsto \:
\exp \, \left( \, \int_S \, B \, \right) \,
\exp \, \left( \, \int_{\partial S} \Lambda \, \right)
\end{displaymath}
As a result, one naively would expect that the integral of $B$
over region $D$ would receive a contribution
\begin{equation}   \label{naive1loop}
\exp \, \left( \, \int^{hx}_x \, \Lambda(g) \: - \:
\int^{gx}_x \Lambda(h) \, \right)
\end{equation}
from the gauge transformations of the $B$ field at the boundaries.
(Relative signs are determined by a choice of orientation.)

However, equation~(\ref{naive1loop}) cannot be the correct answer.
We mentioned earlier that we are free to gauge transform any of the bundles
$T^g$ -- but expression~(\ref{naive1loop}) is not invariant under
gauge transformations of the bundles.

In order to fix~(\ref{naive1loop}) so as to get a gauge-invariant
result, we need to consider the gauge transformations at the corners.
For example, the right end of side~3 in figure~(\ref{fig1loop})
is gauge transformed by $g^* T^h$, whereas the top end of side~4
is gauge transformed by $h^* T^g$, and as noted elsewhere, these
need not be identical.  Phrased differently,
we need to find factors to add to expression~(\ref{naive1loop})
that soak up the gauge transformations of
\begin{displaymath}
\Lambda(g)_{hx} \: - \: \Lambda(g)_x \: - \: \Lambda(h)_{gx}
\: + \: \Lambda(h)_x
\: = \: 
h^* \Lambda(g)_x \: - \: \Lambda(g)_x \: - \: g^* \Lambda(h)_x
\: + \: \Lambda(h)_x
\end{displaymath}
To fix these corner contributions, consider the following expression relating
$\omega^{g,h}$ and the $\Lambda(g)$'s:
\begin{displaymath}
\Lambda(gh) \: = \: \Lambda(h) \: + \: h^* \Lambda(g) \: + \:
d \log \omega^{g,h}
\end{displaymath}
By substracting this expression from the expression involving
$\omega^{h,g}$, we find that
\begin{displaymath}
\left[ \, \Lambda(h) \: + \: h^* \Lambda(g) \, \right]
\: - \: \left[ \, \Lambda(g) \: + \: g^* \Lambda(h) \, \right]
\: = \: d \left[ \, \log \omega^{h,g} \: - \: \log \omega^{g,h} \,
\right]
\end{displaymath}
Using this it is clear that the corrected one-loop phase
correction is given by
\begin{equation}  \label{good1loop}
\left( \omega^{g,h}_x \right) \,
\left( \omega^{h,g}_x \right)^{-1} \,
\exp \, \left( \, \int^{hx}_x \, \Lambda(g) \: - \:
\int^{gx}_x \Lambda(h) \, \right)
\end{equation}
Our derivation leaves an overall constant factor ambiguous; this factor
can be fixed by {\it e.g.} comparing to $B$-field holonomies on
$T^n = {\bf R}^n/{\bf Z}^n$.

Under gauge transformations of the individual bundles $T^g$,
\begin{eqnarray*}
\Lambda(g) & \mapsto & \Lambda(g) \: + \: d \log \phi^g \\
\log \omega^{g,h} & \mapsto &
\log \omega^{g,h} \: + \: \log \phi^{gh} \: - \:
\log \phi^h \: - \: h^* \log \phi^g
\end{eqnarray*}
and it is straightforward to check that equation~(\ref{good1loop})
is indeed invariant under these gauge transformations.

How do the phases in equation~(\ref{good1loop}) compare to the
phases listed in \cite{vafa1}?  It is straightforward to check that
they are the same.  Recall that in order to explicitly recover
representative cocycles, one maps the topologically trivial
bundles $T^g$ to the canonically
trivial bundle and gauge transformations the gauge-trivial connections
$\Lambda(g)$ to the zero connection.  The remaining $\omega^{g_1, g_2}$
are then contant gauge transformations.  If we evaluate 
expression~(\ref{good1loop}) in the described gauge, which we
are free to do since expression~(\ref{good1loop}) is gauge-invariant,
we see 
immediately that the
phases associated to twisted sectors are given by
\begin{equation}    \label{vafat2phase}
\left( \omega^{g,h} \right) \, \left( \omega^{h,g} \right)^{-1}
\end{equation}
which are precisely the phases listed in \cite{vafa1} for a twisted
sector contribution from a pair $(g,h)$.

Since the phase in~(\ref{vafat2phase}) depends only on $g$ and $h$,
and is independent of all other details of the polygon (such as
the location of corners, winding numbers of the sides, and so forth),
it is the same for all contributions to $Z_{(g,h)}$, and so
multiplies $Z_{(g,h)}$ by an overall phase, precisely
as observed in \cite{vafa1}.

In passing, note that the phases~(\ref{vafat2phase}) are invariant
under changing the group cocycles by a group coboundary -- indeed, they must be,
otherwise we could not meaningfully associate phases to elements
of $H^2(\Gamma, U(1))$.

To summarize our progress so far, we have just successfully
derived the twisted sector phases appearing at one-loop
in \cite{vafa1}.

\subsubsection{Invariant analysis}

Next, we shall back up and redo this derivation somewhat more invariantly.
In order to derive these phases, we took the bundles $T^g$ to not only
be topologically trivial, but actually the canonical trivial bundle,
so that the bundle morphisms $\omega^{g_1, g_2}$ would be gauge
transformations, which we used implicitly in writing~(\ref{good1loop}).
These phases can also be described more invariantly, which we shall
now do.

To describe the phases more invariantly, note that the Wilson line
\begin{displaymath}
\int^{gx}_x \, \Lambda(h)
\end{displaymath}
defines a map $T^h_x \rightarrow T^h_{gx} = g^* T^h_x$, so we can think
of this Wilson line as defining an element of 
\begin{displaymath}
\left( T^h_x \right)^{\vee} \, \otimes \, \left( g^* T^h_x \right)
\end{displaymath}
and we can think of the difference of Wilson lines in~(\ref{naive1loop})
as an element of
\begin{equation}    \label{tprod}
\left[ \, \left( T^g_x \right)^{\vee} \otimes \left( h^* T^g_x \right)
\, \right] \: \otimes \:
\left[ \, \left( T^h_x \right)^{\vee} \otimes \left( g^* T^h_x \right)
\, \right]^{\vee}
\end{equation}
which is mapped by the composition of bundle morphisms
$\left( \omega^{h,g} \right)^{-1} \circ \omega^{g,h}$
to a scalar.  So, if we interpret the naive difference of Wilson 
lines~(\ref{naive1loop}) as an element in~(\ref{tprod}),
then we can write the $T^2$ $(g,h)$ twisted sector phase as
\begin{equation}    \label{rgood1loop}
\left( \omega^{h,g} \right)^{-1} \circ
\left( \omega^{g,h} \right) \,
\left[ \, \int^{hx}_x \, \Lambda(g) \: - \:
\int^{gx}_x \, \Lambda(h) \, \right]
\end{equation}
Note the expression~(\ref{rgood1loop}) makes no assumptions concerning
the nature of the bundles or the connections on them, 
unlike expression~(\ref{good1loop}), and is well-defined
under bundle isomorphisms.

\subsubsection{Treatment of the shift orbifold degrees of freedom}

In section~\ref{derivh2}, we argued that in addition to
elements of $H^2(\Gamma, U(1))$, if the covering space $X$ is not
simply connected, or if there are torsion elements in $H^2(X,Z)$,
then one can get additional possible gauge transformations of $B$
fields, beyond those classified by elements of $H^2(\Gamma, U(1))$.
Using the methods of this section, we can now see how such 
gauge transformations would appear in considering orbifold partition
functions.

Suppose $X$ is not simply connected, and one of the bundles $T^g$
defining a gauge transformation of the $B$ field is topologically
trivial, but has a connection
with nontrivial holonomy around some cycle.  From equation~(\ref{good1loop}),
we see that in a $(g,h)$ twisted sector, the phase will no longer
be merely $\omega^{g,h} \left( \omega^{h,g} \right)^{-1}$,
but will also receive a winding-number-dependent contribution.
In other words, instead of merely multiplying the $(g,h)$ partition
function $Z_{(g,h)}$ by a phase, the partition function $Z_{(g,h)}$
will itself be altered.

In other words, a more detailed analysis shows that, in principle,
there can be more degrees of freedom than those found in \cite{vafa1},
which do more than just multiply twisted sector contributions by phases,
but alter those twisted sector contributions.

We will argue in \cite{dtshift} that these degrees of freedom
are not new, but rather are precisely shift orbifolds,
which play an important role in asymmetric orbifolds,
but are less interesting in symmetric orbifolds.
In retrospect, one should have guessed that shift orbifolds
would appear in our analysis -- after all, conventional lore
attributes shift orbifold degrees of freedom to the $B$ field,
in addition to discrete torsion.

\subsection{Two loops}   \label{2loop}

In this section we shall check our calculation of one-loop
phase factors, by repeating the same calculation at the two-loop
level.  In particular, we shall explicitly verify factorization
of the phase factors for two loops into a product of phase factors
for one-loop diagrams.

We shall briefly review factorization, then we shall move on
to the two-loop calculation.

\subsubsection{Review of factorization}

There is an old notion \cite{sw86} that higher-loop string
amplitudes are constrained by one-loop string amplitudes. 
This notion, known as factorization, was used in \cite{vafa1}
to write down the phase factors for twisted sector contributions
to higher-loop partition functions, in terms of the phase factors
for one-loop partition functions.  

\begin{figure}
\centerline{\psfig{file=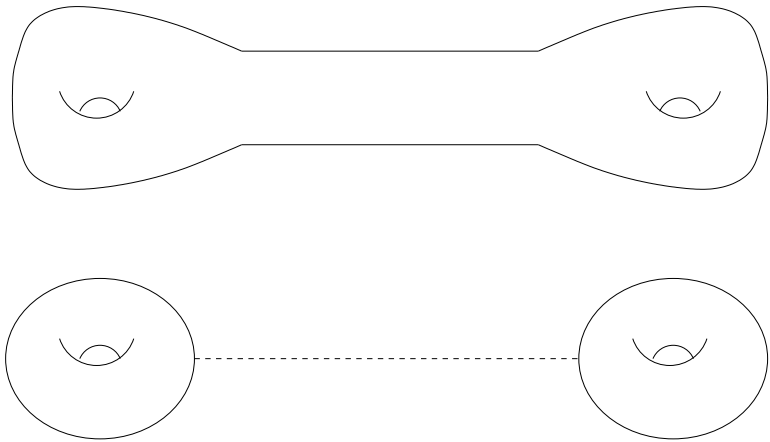,width=4in}}
\caption{  \label{figfact}
Degeneration of a genus two Riemann surface.}
\end{figure}

For two-loop partition functions, the general idea
can be expressed rather briefly.  Since genus two Riemann
surface can degenerate into a pair of genus one Riemann surfaces,
separated by a long thin tube, as sketched in figure~(\ref{figfact}),
and since the twisted sector phase is independent of the moduli
of the Riemann surface,
one finds that the twisted sector phase for a two-loop diagram must be 
a product of
the twisted sector phases for each of the two one-loop diagrams
appearing in the degeneration.

More specifically, if $\epsilon(g_1, h_1; g_2, h_2)$ denotes the
twisted sector phase of a two-loop diagram determined by
$g_1, h_1, g_2, h_2 \in \Gamma$, and this diagram can degenerate
into a product of one-loop diagrams determined by
$(g_1, h_1)$, $(g_2, h_2)$, with twisted sector phases
$\epsilon(g_1, h_1)$ and $\epsilon(g_2, h_2)$, then
\cite{vafa1}:
\begin{equation}    \label{factorization}
\epsilon(g_1, h_1; g_2, h_2) \: = \: \epsilon(g_1, h_1) \,
\epsilon(g_2, h_2)
\end{equation}

In the next subsection we shall verify this factorization condition
explicitly, by calculating the twisted sector phase
$\epsilon(g_1, h_1; g_2, h_2)$ for a two-loop partition function,
and checking that if this can degenerate into a product
of two one-loop contributions, then equation~(\ref{factorization})
holds.

In order to perform the calculation in the next section,
we shall assume $B \equiv 0$, as in the one-loop case,
so that the action on the $B$ field is completely determined
by the gauge transformations (of $B$ fields) distinguishing the
action from the trivial action, and furthermore we shall restrict
to gauge transformations determined by canonically trivial bundles
$T^g$ with gauge-trivial connections $\Lambda(g)$, i.e., those
which correspond to elements of $H^2(\Gamma, U(1))$.

\subsubsection{Two loop calculation}

\begin{figure}  
\centerline{\psfig{file=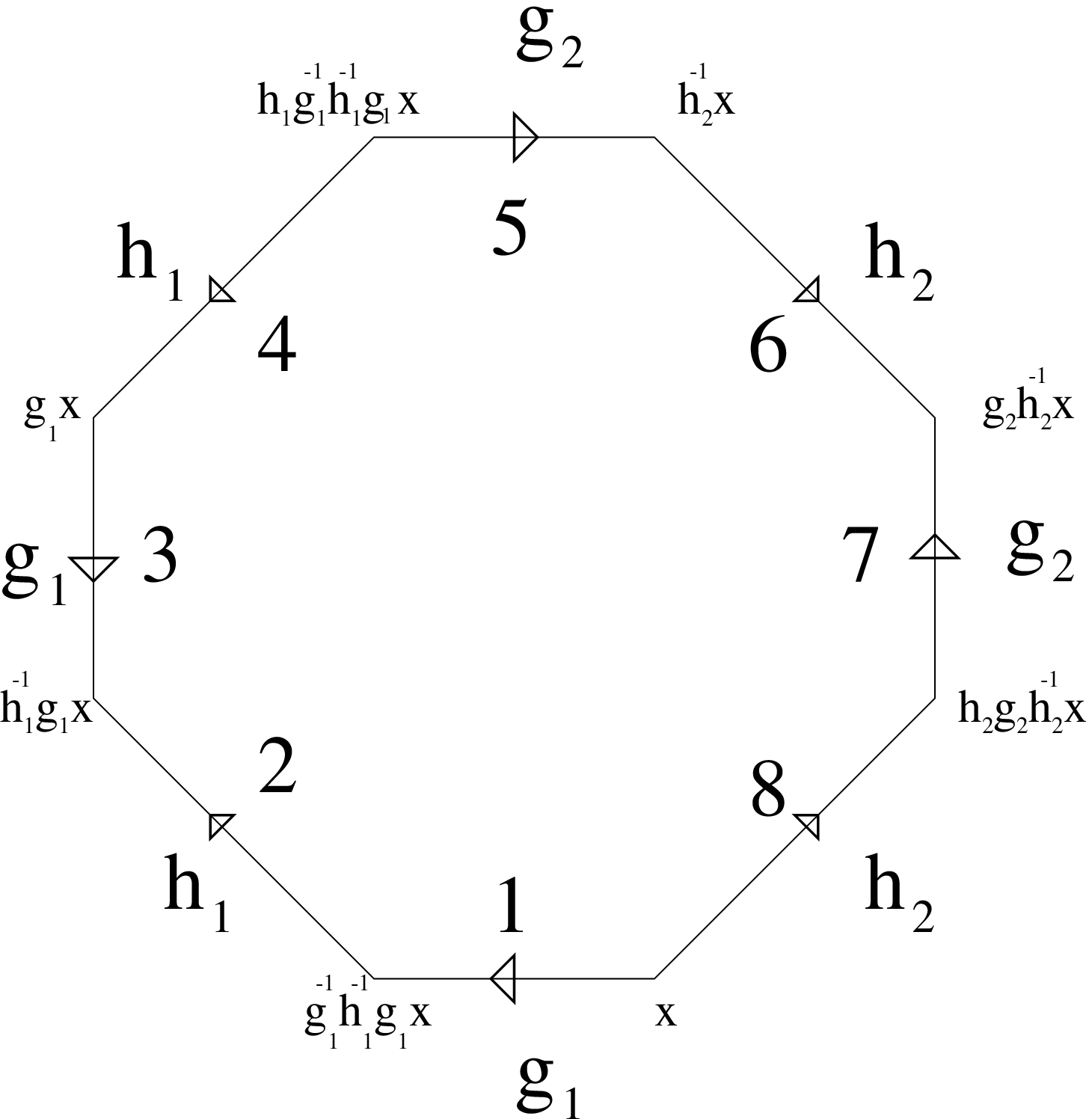,width=3in}}
\caption{  \label{fig2loop}
A twisted sector
contribution to the two-loop partition function.}
\end{figure}

The string orbifold two-loop partition function receives
contributions not only from genus two Riemann surfaces in the covering
space, but also from configurations of strings that form genus two
Riemann surfaces on the quotient space, but only form open polygons
on the covering space, as illustrated in figure~(\ref{fig2loop}).

If two sides of the octagon in figure~(\ref{fig2loop}) are
labelled with the same group element, it indicates that the
two sides are identified under the action of that group element.
For example, sides 1 and 3 are related by $g_1 \in \Gamma$,
with orientations as indicated by the arrows.

In order for the sides of the octagon shown in figure~(\ref{fig2loop})
to close (i.e., in order to have an octagon), we must demand that
\begin{equation}    \label{octdef}
h_1 g^{-1}_1 h^{-1}_1 g_1 \: = \:
g^{-1}_2 h_2 g_2 h^{-1}_2
\end{equation}
In addition, we want to be able to deform this two-loop diagram
into a pair of orbifold one-loop diagrams, determined by the
pairs $(g_1, h_1)$, $(g_2, h_2)$, connected by a long thin
tube.  In order for those one-loop diagrams to be well-defined,
(i.e., in order for the limit described to exist at all),
we must demand that
\begin{displaymath}
g_1 h_1 \: = \: h_1 g_1 \: \: \: \mbox{    and    } \: \: \:
g_2 h_2 \: = \: h_2 g_2
\end{displaymath}
Note that these constraints trivially satisfy equation~(\ref{octdef}).
Also note that these constraints do not imply that every element
of the set $(g_1, h_1, g_2, h_2)$ need commute with every other element 
 -- we have not imposed any sort of commutivity condition relating
$g_1$ and $g_2$, or $h_1$ and $h_2$, for example.

Following the same procedure as for the one-loop case, first note
that the $B$ fields on the sides are related by gluing as
\begin{eqnarray*}
B_3 & = & B_1 \: + \: d \Lambda(g_1) \\
B_4 & = & B_2 \: + \: d \Lambda(h_1) \\
B_7 & = & B_5 \: + \: d \Lambda(g_2) \\
B_8 & = & B_6 \: + \: d \Lambda(h_2)
\end{eqnarray*}
From these identifications,
we naively calculate that the phase defined by orbifold identifications
is 
\begin{equation}    
\exp \, \left( \, \int^{g_1^{-1} h_1^{-1} g_1 x}_x \Lambda(g_1) \: + \:
\int^{h_1^{-1} g_1 x}_{g_1^{-1} h_1^{-1} g_1 x} \Lambda(h_1) 
\: + \:
\int^{h_2^{-1} x}_{h_1 g_1^{-1} h_1 g_1 x} \Lambda(g_2) \: + \:
\int^{g_2 h_2^{-1} x}_{h_2^{-1} x} \Lambda(h_2) \, \right)
\end{equation}
As for the one-loop case, this cannot be the correct result,
simply because it is not invariant under gauge transformations
of the bundles $T^{g_1}$, $T^{h_1}$, $T^{g_2}$, $T^{h_2}$.

As for the one-loop case, we need to determine what to add to
the expression above to correct the phases and fix the corner
contributions.  Specifically, the connections at the endpoints
of the integrals in the naive phase calculation above are given by
\begin{eqnarray*}
\lefteqn{ (g_1^{-1} h_1^{-1} g_1)^* \Lambda(g_1)_x \: - \: \Lambda(g_1)_x
\: + \: (h_1^{-1} g_1)^* \Lambda(h_1)_x \: - \: 
(g_1^{-1} h_1^{-1} g_1)^* \Lambda(h_1)_x } \\
 \: & \: \: & \: + \:
(h_2^{-1})^* \Lambda(g_2)_x \: - \:
(g_2^{-1} h_2 g_2 h_2^{-1} )^* \Lambda(g_2)_x
\: + \: (g_2 h_2^{-1})^* \Lambda(h_2)_x \: - \:
(h_2^{-1})^* \Lambda(h_2)_x
\end{eqnarray*} 
Define $y = g_1^{-1} h_1^{-1} g_1 x$ and
$z = h_2^{-1} x$, then we can rewrite the connections at the
integral endpoints as
\begin{eqnarray*}
\lefteqn{ \Lambda(g_1)_y \: - \: (g_1^{-1} h_1 g_1)^* \Lambda(g_1)_y \: + \:
g_1^* \Lambda(h_1)_y \: - \: \Lambda(h_1)_y } \\
\: & \: \: & \: + \:
\Lambda(g_2)_z \: - \: (g_2^{-1} h_2 g_2)^* \Lambda(g_2)_z \: + \:
g_2^* \Lambda(h_2)_z \: - \: \Lambda(h_2)_z
\end{eqnarray*}

Next, using the relation
\begin{displaymath}
\Lambda(g) \: + \: g^* \Lambda(h) \: - \: \Lambda(h) \: - \:
h^* \Lambda(g) \: = \:
d \left( \, \log \omega^{g,h} \: - \: \log \omega^{h,g} \, \right)
\end{displaymath}
and the assumption that $g_1 h_1 = h_1 g_1$ and $g_2 h_2 =
h_2 g_2$,
we find that the naive expression can be
corrected to give the gauge-invariant phase
\begin{eqnarray*}     
\lefteqn{ \left( \omega^{g_1, h_1}_y \right) \,
\left( \omega^{h_1, g_1}_y \right)^{-1} \,
\left( \omega^{g_2, h_2}_z \right) \,
\left( \omega^{h_2, g_2}_z \right)^{-1} } \\
\: & \: \: & \: \cdot \,
\exp \left( \, \int^{g_1^{-1} h_1^{-1} g_1 x}_x \Lambda(g_1) \: + \:
\int^{h_1^{-1} g_1 x}_{g_1^{-1} h_1^{-1} g_1 x} \Lambda(h_1) \: + \:
\int^{h_2^{-1} x}_{h_1 g_1^{-1} h_1 g_1 x} \Lambda(g_2) \: + \:
\int^{g_2 h_2^{-1} x}_{h_2^{-1} x} \Lambda(h_2)
\, \right)
\end{eqnarray*}

To evaluate the corrected expression above for the two-loop phase,
gauge transform the connections $\Lambda(g)$ to be identically zero,
so the gauge transformations $\omega^{g,h}$ become constant
(assuming the covering space is connected, of course).
Then we find that the correct expression for the two-loop phase is
given by
\begin{equation}
\left( \omega^{g_1, h_1} \right) \,
\left( \omega^{h_1, g_1} \right)^{-1} \,
\left( \omega^{g_2, h_2} \right) \,
\left( \omega^{h_2, g_2} \right)^{-1} 
\end{equation}

This expression is precisely the product of the phases for the
two one-loop contributions determined by $(g_1, h_1)$ and
$(g_2, h_2)$.  Thus, we have explicitly verified factorization
at two loops, by explicitly calculating the twisted sector phase
and noticing, at the end, that it factors, as desired.

Moreover, it is now patently obvious that by repeating the
same calculation at any loop order, we will continue to
find factorization -- so our calculation of twisted sector
phases obeys factorization at all loops.

\section{Derivation of orbifold group actions on D-branes}   
\label{derivdoug}

A description of discrete torsion for D-branes
was proposed by M. Douglas in \cite{doug1,doug2}.  Specifically, he
proposed that turning on discrete torsion in a D-brane orbifold
could be understood as replacing the honest representation
of the orbifold group on the bundle on the D-brane, with
a projective representation.  This description is extremely
natural -- projective representations are classified by
the same group cohomology group as discrete torsion,
namely $H^2(\Gamma, U(1))$ -- and the author further justified
his proposal by, for example, showing how the twisted sector
phases of \cite{vafa1} could be derived from these projective
representations.  The association between discrete torsion and
projective representations was further justified in \cite{gomis},
which gave additional evidence for the relationship.

In this section we shall give a first-principles derivation
of these projective representations proposed in \cite{doug1,doug2}.
We shall also liberally use the conventions, notation,
and general methods of
section~\ref{derivh2}.

In a nutshell, because gauge transformations $B \mapsto B + d \Lambda$
are accompanied by $A \mapsto A + \Lambda$ on a D-brane worldvolume,
in the presence of nontrivial gauge fields, the ``bundle'' on the
worldvolume of the brane is twisted, and one is rapidly led to discover
that orbifold group actions on such twisted ``bundles''
are (appropriately) projectivized.  
Such twisting has been
pointed out explicitly in \cite{freeded}, and has also been used
to describe K-theory in $B$ field backgrounds \cite[section 5.3]{edktheory}.
See also \cite{kapustin} for another recent discussion of this twisting
of D-brane gauge bundles by the $B$ field.
Twisted sheaves have appeared in \cite{caldararu} as part of
an attempt to work out an appropriate version of the
generalized McKay correspondence \cite{mukaimckay} in the presence
of discrete torsion.

In terms of the description of $B$ fields involving stacks
(see for example \cite{dt2}), such twisted sheaves should be
understood as special kinds of sheaves on stacks
\cite{benzvipriv}.  We shall not use such methods here, however.

For simplicity, we shall assume that the $B$ field background
is topologically trivial (i.e., that the curvature $H$ is
zero in integral cohomology).  Our methods are very straightforward
to generalize to $H \neq 0$, but the results are messier, and we see no need
to include the generalization in this paper.

\subsection{General analysis}      \label{dbranact}

Recall that in section~\ref{derivh2} we described $B$ fields
in terms of a collection of data $(B^{\alpha},
A^{\alpha \beta}, h_{\alpha \beta \gamma})$,
where
\begin{eqnarray*}
B^{\alpha} \: - \: B^{\beta} & = & d A^{\alpha \beta} \\
A^{\alpha \beta} \: + \: A^{\beta \gamma} \: + \: A^{\gamma \alpha}
& = & d \log h_{\alpha \beta \gamma} \\
\delta h_{\alpha \beta \gamma} & = & 1
\end{eqnarray*}
As in section~\ref{derivh2}, we shall work on a ``good invariant''
cover $\{ U_{\alpha} \}$, meaning a cover that is well-behaved
with respect to the action of the orbifold group $\Gamma$.

Next, consider the gauge fields\footnote{Unfortunately, we shall
use $A$ to denote both the gauge field on the $N$ D-branes,
as well as part of the \v{C}ech-de Rham cocycle defining the $B$ fields.
These are distinct objects; our notation is unfortunate.
The gauge fields will be associated with single elements of the
cover $\{ U_{\alpha} \}$, and so we shall usually denote the gauge
fields by $A^{\alpha}$.  The \v{C}ech-de Rham cocycle elements
live on intersections $U_{\alpha} \cap U_{\beta} = U_{\alpha \beta}$,
and so we shall denote them by $A^{\alpha \beta}$.  Hopefully
the reader will not become too confused by this.} 
$A$ on a set of $N$ coincident D-branes.
Recall that in D-branes, gauge transformations of
the $B$ field and  
of the gauge fields 
are linked:  $B \mapsto B + d \Lambda$ induces $A \mapsto A + I \Lambda$
(where $I$ denotes the unit matrix, generating the overall $U(1)$
of $U(N)$).

In the present case, this means that since the $B$ fields are only
defined on local charts, the gauge fields $A$ must also only be defined
on local charts in general, with overlaps partially determined by the
same gauge transformations relating the $B$ fields on overlaps \cite{freeded}.
Specifically, to describe the gauge fields on a D-brane
in the presence of $B$ fields as above,
we supplement the data for $B$ fields by data for local $U(N)$ gauge
fields $A^{\alpha}$ as
\begin{eqnarray}
A^{\alpha} \: - \: g_{\alpha \beta} A^{\beta} g^{-1}_{\alpha \beta}
\: - \: d \log g^{-1}_{\alpha \beta} & = & A^{\alpha \beta} I   
\label{aoverlapg} \\
g_{\alpha \beta} \, g_{\beta \gamma} \, g_{\gamma \alpha} \,
& = & h_{\alpha \beta \gamma} I     \label{gandh}
\end{eqnarray}
following \cite{freeded}.  As before, $I$ is the unit matrix,
generating the diagonal $U(1)$ in $U(N)$, $g_{\alpha \beta}$
is an invertible $N \times N$ matrix that would describe transition
functions for the bundle if the $B$ field were completely trivial,
and $A^{\alpha}$ is a local $U(N)$ gauge field on the D-brane.
Also note we are using $d \log g^{-1}_{\alpha \beta}$ as shorthand for 
$g_{\alpha \beta} d g^{-1}_{\alpha \beta}$.

Now, ordinarily to describe how an orbifold group $\Gamma$ acts
on a bundle, we would demand
\begin{displaymath}
g^* g_{\alpha \beta} \: = \: \left( \gamma^g_{\alpha} \right) \,
\left( g_{\alpha \beta} \right) \, \left( \gamma^g_{\beta} \right)^{-1}
\end{displaymath}
for some $N \times N$ matrices $\gamma^g_{\alpha}$.
However, because of equation~(\ref{gandh}), we have to be more careful.
From studying the pullback of equation~(\ref{gandh}) by an element
$g \in \Gamma$, we find
\begin{eqnarray*}
g^* \left[ \, \left( g_{\alpha \beta} \right) \,
\left( g_{\beta \gamma} \right) \,
\left( g_{\gamma \alpha} \right)  \, \right] & = &
\left( g^* h_{\alpha \beta \gamma} \right) \, I \\
 & = & h_{\alpha \beta \gamma} \, \nu^g_{\alpha \beta} \,
\nu^g_{\beta \gamma} \, \nu^g_{\gamma \alpha} \, I
\end{eqnarray*}
where the $\nu^g$ are \v{C}ech cochains that appeared in
section~\ref{derivh2} in describing the orbifold group action on
the $B$ fields themselves.
From the above, we see that the most general expression for
$g^* g_{\alpha \beta}$ is given by
\begin{equation}
g^* g_{\alpha \beta} \: = \:
\left[ \, \left( \gamma^g_{\alpha} \right) \,
\left( g_{\alpha \beta} \right) \,
\left( \gamma^g_{\beta} \right)^{-1} \, \right] \,
\left[ \, \left( \nu^g_{\alpha \beta} \right) \, I \, \right]
\end{equation}
where the $\gamma^g_{\alpha}$ are some locally-defined $U(N)$ adjoints.
In short, we find that our naive expression for $g^* g_{\alpha \beta}$
is twisted.

From expressing $(g_1 g_2)^* g_{\alpha \beta}$ in two different ways
(following the general self-consistency bootstrap outlined
in section~\ref{derivh2}), we find a constraint on the $\gamma^g$:
\begin{equation}   \label{gammaproj}
\left( h^{g_1, g_2}_{\alpha} \right) \, \left( \gamma^{g_1 g_2}_{\alpha}
\right) \: = \:
\left( g_2^* \gamma^{g_1}_{\alpha} \right) \,
\left( \gamma^{g_2}_{\alpha} \right)
\end{equation}

In fact, equation~(\ref{gammaproj}) above already tells us that ordinary lifts
of orbifold group actions must be replaced by projective lifts,
as hypothesized in \cite{doug1,doug2}, but we shall finish working out 
the orbifold group action on the D-brane gauge fields before 
emphasizing this point in detail.

Next, write
\begin{displaymath}
g^* A^{\alpha} \: = \: u(g)^{\alpha} \, A^{\alpha} \,
\left( u(g)^{\alpha} \right)^{-1} \: + \:
A(g)^{\alpha}
\end{displaymath}
for some $U(N)$ adjoints $u(g)^{\alpha}$, $A(g)^{\alpha}$.
This equation is not a constraint -- it is general enough
to describe any possible $g^* A^{\alpha}$, by varying $u(g)^{\alpha}$
and $A(g)^{\alpha}$.  We are merely writing
$g^* A^{\alpha}$ in a form that will yield more understandable results,
following the self-consistent bootstrap of section~\ref{derivh2}.
By pulling back both sides of equation~(\ref{aoverlapg}) by
$g \in \Gamma$ and examining the result, we can determine
both $u(g)^{\alpha}$ and $A(g)^{\alpha}$:
\begin{eqnarray}
u(g)^{\alpha} & = & \gamma^g_{\alpha} \\
A(g)^{\alpha} & = & \left( \gamma^g_{\alpha} \right) \,
d \left( \gamma^g_{\alpha} \right)^{-1} \: + \:
I \Lambda(g)^{\alpha}
\end{eqnarray}
where $\Lambda(g)^{\alpha}$ is a set of local one-forms
that appeared in section~\ref{derivh2} in defining the action of the
orbifold group on the $B$ fields themselves.

To summarize, we have found that the orbifold group action
on the $U(N)$ gauge fields on the worldvolume of a D-brane is described
by
\begin{eqnarray*}
g^* A^{\alpha} & = & \left( \gamma^g_{\alpha} \right) \,
A^{\alpha} \, \left( \gamma^g_{\alpha} \right)^{-1} \: + \:
\left( \gamma^g_{\alpha} \right) \,
d \left( \gamma^g_{\alpha} \right)^{-1} \: + \:
I \Lambda(g)^{\alpha} \\
g^* g_{\alpha \beta} & = & \left( \nu^g_{\alpha \beta} \right) \,
\left[ \, \left( \gamma^g_{\alpha} \right) \,
\left( g_{\alpha \beta} \right) \,
\left( \gamma^g_{\beta} \right)^{-1} \, \right] \\
\left( h^{g_1, g_2}_{\alpha} \right) \,
\left( \gamma^{g_1 g_2}_{\alpha} \right) & = &
\left( g_2^* \gamma^{g_1}_{\alpha} \right) \,
\left( \gamma^{g_2}_{\alpha} \right)
\end{eqnarray*}
where $\Lambda(g)$ and $\nu^g_{\alpha \beta}$ were defined in
section~\ref{derivh2}, in defining the action of the orbifold group
on the $B$ fields themselves.

As we shall work out explicitly below,
the data above describes the projectivized orbifold group
actions described in \cite{doug1,doug2}, but in 
considerably greater generality.

\subsection{Explicit comparison to results of M. Douglas}

The result above is considerably more general than that appearing
in \cite{doug1,doug2}.  For purposes of comparison, let us take
a moment to specialize to their circumstances.  The papers
\cite{doug1,doug2} considered orbifolds of D-branes in backgrounds
with vanishing $B$ fields, where the D-branes had support on some
flat space ${\bf R}^n$, and had a topologically trivial bundle
on their worldvolume.  Since the $B$ fields are completely trivial,
we can take $B^{\alpha} = 0$, $A^{\alpha \beta} = 0$, and $h_{\alpha
\beta \gamma} = 1$.

Since the $B$ fields are completely trivial, 
it is meaningful to speak of an honest bundle on the worldvolume
of the D-brane -- transition functions close on overlaps, not just
up to a phase.  Another assumption made in \cite{doug1,doug2}
is that this bundle on the D-brane worldvolume is topologically
trivial.  Since it is topologically trivial, we can take the
$g_{\alpha \beta} = 1$, and replace the locally-defined
gauge fields $A^{\alpha}$ with a single global $U(N)$ gauge field
$A$.  In addition, \cite{doug1,doug2} make the further assumption
that the gauge field $A$ is constant, so in defining the action of
the orbifold group, it suffices to assume $\gamma^g$ is constant.

Since the $B$ fields are trivial (in fact, vanishing), we can
describe any orbifold group action on the $B$ fields by specifying
the difference between that action and the canonical trivial action.
In other words, since the $B$ field background is trivial,
we can talk about orbifold group actions specified by elements
of $H^2(\Gamma, U(1))$ (whereas, in general, only the difference
between two orbifold group actions could be described in this fashion).

An orbifold group action specified by an element of $H^2(\Gamma, U(1))$
can be specified by a set of topologically trivial bundles $T^g$
with gauge-trivial connection $\Lambda(g)$, and bundle maps $\omega^{g_1, g_2}$.
As noted earlier, without loss of generality we can take the
bundles $T^g$ to all be the canonical trivial bundle, and the
connections $\Lambda(g)$ to all be identically zero, so the
$\omega^{g_1, g_2}$ become constant gauge transformations of the trivial
bundle, satisfying the group 2-cocycle condition.

In terms of the data describing the orbifold group action on the D-brane
worldvolume gauge fields, this means we can take the $\nu^g_{\alpha \beta} = 1$,
and the $h^{g_1, g_2}_{\alpha}$ to be functions, i.e.,
\begin{displaymath}
h^{g_1, g_2}_{\alpha} \: = \: h^{g_1, g_2}_{\beta} \mbox{  on } U_{\alpha}
\cap U_{\beta}
\end{displaymath}
and in fact constant functions,
satisfying the group 2-cocycle condition 
\begin{displaymath}
\left( h^{g_1, g_2 g_3} \right) \,
\left( h^{g_2, g_3} \right) \: = \:
\left( h^{g_1, g_2} \right) \,
\left( h^{g_1 g_2, g_3} \right)
\end{displaymath}
Since we assumed the $B$ fields to be completely trivial,
and we are describing all lifts in terms of gauge transformations
combined with the canonical trivial lift (that exists in this case),
the functions $h^{g,h}$ coincide with constant functions we
have labelled $\omega^{g,h}$ elsewhere -- in other words,
the 2-cocycles $h^{g,h}$ are the same cocycles as those describing
the relevant element of $H^2(\Gamma, U(1))$.

We can now rewrite the action of the orbifold group on the D-brane
gauge fields, in this special case, as
\begin{eqnarray*}
g^* A & = & \left( \gamma^g \right) \, A \, \left( \gamma^g \right)^{-1} \\
\left( h^{g_1, g_2} \right) \, \left( \gamma^{g_1 g_2} \right)
& = &
\left( \gamma^{g_1} \right) \,
\left( \gamma^{g_2} \right)
\end{eqnarray*}
which is precisely the projectivized orbifold group action described
in \cite{doug1,doug2}.

In recovering the form of the orbifold group action described
in \cite{doug1,doug2}, we made the same simplifying assumptions
that appeared in \cite{doug1,doug2} -- namely, that the D-branes
reside on ${\bf R}^n$, and the bundle on the D-brane worldvolume
is topologically trivial, with constant gauge field $A$.
However, our results from the previous subsection
apply in much more general circumstances
than these.  Put another way, we have just shown how to derive
the description of discrete torsion for D-branes given in
\cite{doug1,doug2}, but our results apply in far greater
generality than that used in \cite{doug1,doug2}.

\subsection{Notes on shift orbifolds}

In section~\ref{derivh2} we pointed out that there are
additional orbifold group actions on $B$ fields,
in addition to those classified by elements of $H^2(\Gamma, U(1))$,
which correspond to the so-called shift orbifolds \cite{dtshift}.
As noted in section~\ref{twphase},
these new actions do considerably more than just multiply
twisted sector partition functions $Z_{(g,h)}$ by a phase;
they multiply individual contributions to $Z_{(g,h)}$ by
distinct phases, and so they deeply alter $Z_{(g,h)}$.

How do these extra orbifold group actions appear on D-branes?
The answer is implicit in the orbifold group action 
given at the end of section~\ref{dbranact} above.
Our results on D-brane actions made no assumptions regarding
the form of the orbifold group action on the $B$ fields.
In the special case of orbifold group actions on trivial $B$
fields arising from elements of $H^2(\Gamma, U(1))$, we 
re-derived the results of \cite{doug1,doug2};
but our methods apply in general.
We will describe the resulting group actions on D-branes in
more detail in \cite{dtshift}.

\section{Notes on Vafa-Witten}    \label{derivvw}

In the paper \cite{vafaed} C. Vafa and E. Witten analyzed
the interrelationship between discrete torsion and Calabi-Yau moduli.
Specifically, they considered deformations of the orbifolds
$T^6 / \left( {\bf Z}_2 \times {\bf Z}_2 \right)$ and
$T^6 / \left( {\bf Z}_3 \times {\bf Z}_3 \right)$ both with
and without discrete torsion.  In both cases, turning on discrete torsion
had the effect of removing most elements of $H^{1,1}_{orb}$ from the
massless spectrum, while adding new elements to $H^{2,1}_{orb}$.
In both cases they construct certain\footnote{For a more comprehensive
description of Calabi-Yau deformations and resolutions of
such orbifolds than was provided in \cite{vafaed},
see \cite[section 6]{joyce}.} families of Calabi-Yau deformations
of the orbifolds,
in which the elements of $H^{2,1}_{orb}$ form a subset
of the possible polynomial deformations.  Furthermore, they argue that
the ``allowed'' complex structure deformations (those linked to
elements of $H^{2,1}_{orb}$) can not fully resolve the space, but
always leave isolated singularities.  Finally, they attempted
to make sense out of these results by conjecturing the existence
of some sort of ``discrete torsion for conifolds,'' the isolated
singularities left after desingularizing as much as possible.

How can one understand this behavior, in terms of our picture of
discrete torsion?
So far in this paper we have described discrete torsion as
the action of the orbifold group on the $B$ fields.  
In order to understand reference \cite{vafaed} we need to understand
quotients of $B$ fields, directly on singular spaces, and their
behavior under deformation.

As always, there are many close analogues to properties of
orbifold Wilson lines.
We begin in section~\ref{owlvw} by describing the analogues
of \cite{vafaed} in orbifold Wilson lines, where the analysis is vastly
simpler.  Then, in section~\ref{vwoutline} we outline, in general terms,
how the results of \cite{vafaed} can be understood in the framework
we have presented, as (as always) an analogue for $B$ fields of
behavior of $U(1)$ gauge fields.

We shall not attempt to give a detailed first-principles
derivation of the results in \cite{vafaed}, but rather shall only
outline general ideas.  Detailed derivations are deferred to later work.

\subsection{Analogue for orbifold Wilson lines}   \label{owlvw}

In this paper we have often discussed orbifold Wilson lines
on smooth covering spaces.  Understanding orbifold Wilson lines on 
singular quotient spaces is considerably more subtle.

To begin, consider \cite[chapter 14]{duistermaat}
a trivial rank $n$ complex vector bundle on ${\bf C}^2$.
Let ${\bf Z}_2$ act on ${\bf C}^2$ in the standard fashion,
and combine the ${\bf Z}_2$ action on the base with a gauge transformation
that maps fibers to minus themselves, i.e.,
\begin{displaymath}
( z_1, z_2, \cdots, z_n) \: \mapsto ( -z_1, -z_2, \cdots, -z_n)
\end{displaymath}
Now, consider the quotient space.
Over ${\bf C}^2/{\bf Z}_2$, we have some fibration, the quotient of
the total space of the trivial ${\bf C}^n$ vector bundle.
Specifically, over smooth points on ${\bf C}_2/{\bf Z}_2$,
we recover the original ${\bf C}^n$ fiber, but over the singularity
on ${\bf C}^2/{\bf Z}_2$, the fiber is ${\bf C}^n/{\bf Z}_2$.

In other words, the quotient of a bundle, is not a bundle in general.
We can understand this matter in somewhat less technical terms also.
Consider quotienting ${\bf C}^2$ as above.
Consider Wilson loops enclosing the origin in ${\bf C}^2/{\bf Z}_2$.
At least for those loops descending from open strings on ${\bf C}^2$
with ends identified by the ${\bf Z}_2$, there will be a nontrivial
holonomy about the loop.  However, as we contract the loop to the
origin, the holonomy remains nonzero -- so something unusual must
be going on at the origin.  Since we started with a flat connection
on the cover, the connection must be flat on the quotient space
(at least, on the smooth part of the quotient).
Since ${\bf C}^2/{\bf Z}_2$ is contractible,
if there were an honest everywhere flat bundle on the space,
any Wilson loop would be forced to be zero.  Since there are
nonzero Wilson loops, we are forced to conclude that our
``bundle'' must be behaving badly at the origin. 

In order to understand orbifold Wilson lines on singular quotient
spaces, one must work with more general objects than mere bundles.
On complex surfaces, the relevant objects turn out to be 
reflexive, non-locally-free
sheaves.  On a smooth variety, a reflexive sheaf will be locally
free up to complex codimension three, but on a singular variety,
a reflexive sheaf can be locally free at lower codimension, provided
the failure of local freedom occurs over the singularities.
(As reflexive sheaves are not used very widely in the physics
literature, we have included an appendix giving general background
information on reflexive sheaves, as well as an example -- a derivation of 
a reflexive rank 1 sheaf appearing as a ${\bf Z}_2$ quotient
of ${\bf C}^2$ with nontrivial orbifold Wilson lines on the trivial
line bundle over ${\bf C}^2$.)

If someone did not know about reflexive sheaves, and tried to
understand the objects living on singular spaces, they would
probably try to think of them as some analogue of orbifold
Wilson lines.  For example, there exist\footnote{These exist
mathematically, but whether they are relevant for physics is
unknown.} reflexive non-locally-free
sheaves at conifold singularities.  Someone not acquainted with
reflexive sheaves might try to label such objects
``orbifold Wilson lines for conifolds.''

What happens to these reflexive sheaves when the space is resolved?
On a complex surface, a reflexive sheaf will lift to a locally-free
sheaf, i.e., a bundle, on the resolved space.  The new bundle
will typically have nonzero curvature (nonzero $c_1$) associated
with the exceptional divisor of the resolution.  This can be
seen directly in the ADHM/ALE construction, and incidentally forms
one way of understanding\footnote{The classical McKay correspondence
is a map from representations of $\Gamma$, a finite subgroup of $SU(2)$,
to elements of the degree 2 cohomology of $\widetilde{ {\bf C}^2/\Gamma }$.
Here, we see that directly -- a representation of $\Gamma$ defines
an action of the orbifold group on the trivial bundle on ${\bf C}^2$,
and the corresponding element of degree 2 cohomology is $c_1$
of the bundle appearing as a lift of the reflexive sheaf on the
quotient ${\bf C}^2/\Gamma$.  This way of thinking about McKay is
essentially due to \cite{av,gsv}.} the classical McKay correspondence.

One can also understand this resolution in less technical terms.
Consider our previous example of a quotiented bundle over 
${\bf C}^2/{\bf Z}_2$.
We remarked earlier that we have nonzero Wilson loops, when we
naively would have expected all Wilson loops to be zero.  
Now, when we blow up the singularity, we can lift the Wilson loops
to the cover.  If we were to have an everywhere flat bundle on the
resolution, then since $\pi_1 = 0$ and $H^2({\bf Z})$ has no torsion,
no Wilson loop can be nonzero -- a contradiction.
So, in order to avoid contradicting the existence of nonzero Wilson loops,
we are forced to conclude that there must be nonzero curvature someplace,
and as our connection was flat away from the singularity, the curvature
must live on the exceptional divisor of the resolution.
Phrased more naively, quotients of bundles will have nonzero
curvature concentrated at curvature singularities if one turns on
orbifold Wilson lines.



\subsection{$B$ fields on singular spaces}   \label{vwoutline}

We shall not attempt in this paper to give a thorough derivation of the results
of \cite{vafaed}, but in general terms, 
their results should now seem much more natural.

Just as a quotient of a bundle need not be a bundle,
the quotient of a 1-gerbe (a formal structure, analogous to bundles,
for which $B$ fields are connections) need not be a 1-gerbe.
Instead, on a singular variety, one would merely expect
to have algebraic stacks (analogues of sheaves) which fail to
be gerbes over certain quotient singularities.

For example, the ``discrete torsion for conifolds'' observed
by the authors of \cite{vafaed} surely corresponds to some
algebraic stack on a conifold singularity that fails to be
a gerbe locally over the singularity, a precise analogue of
the reflexive sheaves on conifolds that we labelled
``orbifold Wilson lines for conifolds.''

We can justify this claim in elementary terms as follows.  Recall in studying
our example on ${\bf C}^2/{\bf Z}_2$, we concluded that because
of the existence of nonzero Wilson loops on the quotient space,
the resolved space must have nonzero curvature supported at
the exceptional divisor.  Here, we have a very similar situation.
Here we have nonzero Wilson surfaces (i.e, $\exp \left( \int B \right)$)
over Riemann surfaces in the quotient spaces ${\bf C}^3/ \left( 
{\bf Z}_n \times {\bf Z}_n \right)$, as measured by the twisted sector
phases worked out in section~\ref{twphase}.  Just as for orbifold Wilson
lines, this is only sensible here because the spaces have singularities.
If we resolve the singularities, in order to get a consistent picture
we must generate nonzero curvature.

Now, in very general terms, there are two general ways to resolve
singularities: 
\begin{enumerate}
\item We can deform the K\"ahler structure, by blowing up or
making a small resolution.  
\item We can deform the complex structure.
\end{enumerate}

In the first case, although we mentioned that we need nonzero curvature
in order to make sense out of the Wilson surfaces present on the quotient,
in general one does not generate any natural 3-cycles where
the curvature $H$ could be supported, since blowups and
small resolutions add even-dimensional cycles.
Thus, we are naturally led to the conclusion that, in general terms,
resolutions of this class must be obstructed -- we need to turn on
curvature $H$ somewhere to account for the nonzero Wilson surfaces,
but typically one does not have any natural options to do so.
Indeed, in \cite{vafaed} it was noted that nontrivial K\"ahler deformations
are removed from the massless spectrum.

In the second case, one can often get 3-cycles.  For example,
if we smooth a conifold singularity by deforming the complex structure,
then we are typically led to a new 3-cycle.  So, in this case
one expects to typically be able to resolve the space without
contradiction, and generate nonzero curvature $H$ on new 3-cycles.

Now, nonzero curvature $H$ (in de Rham cohomology) on a Calabi-Yau
is inconsistent with supersymmetry -- if a deformation involves
turning on $H$, then that deformation will break supersymmetry,
and so that deformation is lifted from the moduli space of 
supersymmetric vacua.  Indeed, in \cite{vafaed} it was found that
complex structure deformations which completely
resolved the space, and left no singularity, were also obstructed. 


This explanation of how Calabi-Yau moduli of type $H^{2,1}$
could be obstructed 
was suggested in \cite[section 2.3]{vafaed} as
an explanation of their results, though they had no idea
why $H$ should be generated. 
In the present context, this is naturally understood.

So, we see that if we try to deform an orbifold with
nonzero discrete torsion (which implies nonzero Wilson surfaces
on the quotient), then logical consistency and the demands of
supersymmetry place strong constraints on the structure
of the moduli space of vacua, which in general terms is 
consistent with \cite{vafaed}.

As mentioned earlier, we shall not attempt a thorough derivation
of the results of \cite{vafaed} in this paper, but rather
shall defer such detailed considerations to future work.
For the purposes of this paper, we are content to merely
outline how the results of \cite{vafaed} naturally
fit into the general picture we have described.

\section{Notes on local orbifolds}    \label{locorb}

A point usually glossed over in physics discussions of
orbifolds is the possibility of local orbifold degrees of
freedom, as distinct from global orbifold degrees of freedom.
We shall take a moment to very briefly discuss local orbifolds and discrete
torsion, before concluding.

What are local orbifold degrees of freedom?
Consider forming a global orbifold, say, $T^4/{\bf Z}_2$.
In forming the global orbifold, if we are doing physics,
then we usually get to choose some number of degrees
of freedom -- orbifold Wilson lines and discrete torsion, for example.
Now, we can also examine each singularity locally.
In the present case, each singularity is locally ${\bf C}^2/{\bf Z}_2$.
If we forget about the global structure of the orbifold and
just work locally, then one is led to associate orbifold
degrees of freedom to the singularities locally.
We refer to such local degrees of freedom as
local orbifold degrees of freedom.

How do local orbifold degrees of freedom differ from
global orbifold degrees of freedom?
For simplicity, consider a $U(1)$ bundle with connection
on $T^4$, and quotient by ${\bf Z}_2$.  If we turn on 
the nontrivial orbifold $U(1)$ Wilson line, then we find that
each ${\bf C}^2/{\bf Z}_2$ has a nontrivial ``twist'' (technically,
locally we have a reflexive non-locally-free rank 1 sheaf).
So, the global orbifold gives rise to either a twist at
every singularity, or no twists at any singularity.

However, there are additional options present if we consider
the orbifold to be a local orbifold.  We can consistently
twist at groups of eight of the sixteen singularities
(technically, in addition to reflexive non-locally-free sheaves
that fail to be locally free at every singularity or nowhere,
there are also reflexive non-locally-free sheaves that fail
to be locally free at some but not all singularities).
(See \cite[section VIII.5]{bpvdev} or \cite{paul96} for the
mathematical result.)  

One can now ask:  does discrete torsion have local orbifold
degrees of freedom?  Previously, when discrete torsion was 
known merely as some mysterious discrete degree of freedom appearing
in orbifolds, the answer was not known.  

Now that we have a purely mathematical understanding of discrete
torsion, we can address this matter.  In general terms, it is now
clear that there should
exist local orbifold degrees of freedom for discrete
torsion.  We shall not attempt any counting of such local orbifold
degrees of freedom here -- rather, our purpose was merely to
emphasize the important but usually neglected point that
local orbifold degrees of freedom exist.

\section{Conclusions}

In this paper we have given a complete geometric explanation
of discrete torsion, as the choice of orbifold group action on the
$B$ fields.  Specifically, we have shown how the group
cohomology group $H^2(\Gamma, U(1))$ arises, derived the phases
associated with twisted sector contributions to string loop
partition functions, derived M.~Douglas's description \cite{doug1,doug2}
of orbifold
group actions on D-branes as projective representations of the
orbifold group, and outlined how the results of Vafa-Witten
\cite{vafaed} fit into this general framework.
We have also briefly discussed shift orbifolds, which
are degrees of freedom associated
with the $B$ field, beyond those classified by $H^2(\Gamma, U(1))$,
and explained how these appear physically in terms of twisted sector
contributions to partition functions and in terms of D-brane actions.

Nowhere in all this did we assume that the orbifold group
$\Gamma$ acts freely; nor do we assume $\Gamma$ is abelian.
We do not even assume that the curvature of the $B$ fields vanishes.
Our results hold in generality.

To put a different spin on these matters, we have given a completely
geometric description of discrete torsion.  {\it A priori},
discrete torsion has nothing to do with string theory.
Discrete torsion is a property of defining orbifold group
actions on $B$ fields, and can be understood in a purely
mathematical context, without any reference to string theory.
Now, at the end of the day, we can calculate twisted sector
contributions to partition functions, as well as check M.~Douglas's
proposed orbifold group action on D-branes, so we can certainly
derive physical results.  However, it should be emphasized, 
discrete torsion is not some special, ``inherently stringy''
property of string theory or conformal field theory, but rather
has a straightforward and purely mathematical understanding.

\section{Acknowledgements}

We would like to thank P.~Aspinwall, D.~Ben-Zvi, 
A.~Caldararu, A.~Knutson,
D.~Morrison, and R.~Plesser for useful conversations.

\appendix

\section{Reflexive sheaves on quotient spaces}   \label{reflex}

In section~\ref{owlvw}, we mentioned that when orbifolding a space
with a bundle, the resulting object living on the quotient space
need not be a bundle, and in the context of complex algebraic geometry,
will be a reflexive sheaf.  In this appendix we elaborate on these
remarks, as reflexive sheaves are not commonly understood in the
physics literature.  We begin with a short overview of reflexive
sheaves, and then explicitly derive the reflexive sheaf appearing
in the specific example of the quotient ${\bf C}^2/{\bf Z}_2$.

\subsection{Technical notes on reflexive sheaves}    \label{reflexap}

In this appendix we shall give some technical notes on
reflexive sheaves.

A reflexive sheaf is a sheaf ${\cal E}$ which is isomorphic to
its bidual:  ${\cal E} \cong {\cal E}^{\vee \vee}$.
Reflexive sheaves include locally free sheaves (i.e., bundles)
as a special subclass.

On smooth varieties, reflexive sheaves are locally free up to 
(complex) codimension three.  Thus, for example, on a smooth
surface, all reflexive sheaves are locally free.
Also, on a smooth variety, all reflexive rank 1 sheaves are locally free.

On singular varieties, both of the statements above fail.
On a singular surface, one can have reflexive non-locally-free sheaves;
the sheaves fail to be locally free over the singularities.
One can also have reflexive non-locally-free rank 1 sheaves on
singular varieties.

On Noetherian normal varieties (i.e., most of the varieties that
physicists are likely to encounter in practice), we can describe
reflexive rank 1 sheaves in terms of divisors, just as line bundles
on smooth varieties
can be described in terms of divisors.
Roughly speaking, a general divisor on a Noetherian normal variety
is known as a ``Weil'' divisor, whereas a divisor that defines,
not just a reflexive rank 1 sheaf, but a line bundle, is known as
a ``Cartier'' divisor\footnote{Experts will note we are being quite
sloppy--our description of Cartier divisors really corresponds to the
image of Cartier divisors in the space of Weil divisors.
Our abbreviated description will suffice for the purposes of
this appendix.}.

For example, on the singular affine space ${\bf C}^2/{\bf Z}_2$,
there is precisely one (equivalence class of) Weil divisor that
is not Cartier.
If we write 
\begin{displaymath}
{\bf C}^2 / {\bf Z}_2 \: = \: \mbox{Spec } {\bf C}[x,y,z]/(xy-z^2)
\end{displaymath}
then the non-Cartier Weil divisor is $\{ x = z = 0 \}$.
We shall denote this divisor by $D$.
Then, the associated reflexive sheaf ${\cal O}(D)$ (and also
${\cal O}(-D)$) is not locally free.

Although $D$ is not Cartier, it can be shown that the divisor $2D$ is Cartier,
so for example, the reflexive sheaves ${\cal O}(-D)$ and ${\cal O}(D)$
are related by tensoring with a (trivial) line bundle, and so are in
the same equivalence class of Weil divisors.

Similarly, the divisor $\{ y = z = 0 \}$ is not Cartier, but it is related
to $D$ by a Cartier divisor, and so lies in the same equivalence class
of Weil divisors.


Reflexive rank 1 sheaves can be somewhat more subtle than line
bundles.  For example, if $D_1$ and $D_2$ are a pair of Cartier
divisors, then as everyone knows,
\begin{displaymath}
{\cal O}(D_1) \otimes {\cal O}(D_2) \: = \: {\cal O}(D_1 + D_2)
\end{displaymath}
In other words, the tensor products of the line bundles associated
to (Cartier) divisors $D_1$, $D_2$ is the line bundle associated to
(Cartier) divisor $D_1 + D_2$.
However, if $D_1$ and $D_2$ are not both Cartier divisors,
then this relation need not hold \cite[p. 283]{reid}.
In other words, if $D_1$ and $D_2$ are two Weil divisors,
then in general the reflexive rank 1 sheaves ${\cal O}(D_1)$,
${\cal O}(D_2)$ are not related to ${\cal O}(D_1 + D_2)$ as above:
\begin{displaymath}
{\cal O}(D_1) \otimes {\cal O}(D_2) \: \neq \: {\cal O}(D_1 + D_2)
\end{displaymath}
The essential difficulty is that the tensor product of two reflexive
sheaves need not be reflexive -- the tensor product can contain
torsion, for example.
We can fix this problem by bidualizing the left-hand side of the
expression above.  In other words, a statement that is true for
all Weil divisors $D_1$, $D_2$, not just Cartier divisors, is
\begin{displaymath}
\left( {\cal O}(D_1) \otimes {\cal O}(D_2) \right)^{\vee \vee}
\: = \: {\cal O}(D_1 + D_2)
\end{displaymath}

A more detailed discussion of the relationship of Weil versus
Cartier divisors, in the context of toric varieties, can be
found in \cite{sotvfp}.

\subsection{Reflexive sheaves and orbifold Wilson lines}  \label{quotmod}

In the text we claimed that one can directly check that possible
quotients of the structure (trivial rank one) sheaf on ${\bf C}^2$
are precisely the possible reflexive sheaves on ${\bf C}^2/\Gamma$.
Here we shall work this out in detail for the case $\Gamma = {\bf Z}_2$.

Let $R$ denote the ring ${\bf C}[x,y]$, and let $R^{\Gamma}$ denote
the ring of $\Gamma$-invariants, namely
\begin{displaymath}
R^{\Gamma} \: = \: {\bf C}[x^2,y^2,xy]
\end{displaymath}
Let $M$ denote the $R$-module defining the structure sheaf.
In other words, $M = R$.  (The distinction between the module $M$
and the ring $R$ shall be important when discussing nontrivial lifts
of $\Gamma$.)

First, consider the case of the trivial lift of the $\Gamma = {\bf Z}_2$
action from ${\bf C}^2$ to the structure sheaf.
Then the module $M^{\Gamma}$ of $\Gamma$-invariants is isomorphic
(as an $R^{\Gamma}$-module) to the trivial $R^{\Gamma}$-module,
namely
\begin{displaymath}
M^{\Gamma} \: = \: {\bf C}[x^2,y^2,xy]
\end{displaymath}
In this case, $M^{\Gamma}$ defines the structure sheaf on the
quotient space ${\bf C}^2/\Gamma$.  This sheaf is locally free.

Now, consider the nontrivial lift of the $\Gamma = {\bf Z}_2$ action
from ${\bf C}^2$ to the structure sheaf on ${\bf C}^2$.
In order to describe the action of this lift on the module $M$,
view $M$ as a freely generated $R$-module, and let the (single)
generator of $M$ be denoted by $\alpha$.
For the trivial lift, we implicitly assumed that $\Gamma$ mapped
$\alpha \mapsto \alpha$.  For the nontrivial lift,
we take $\alpha \mapsto - \alpha$.
The $R^{\Gamma}$-module of $\Gamma$ invariants of $M$ is now
a module with two generators, namely $x \alpha$ and
$y \alpha$, and one relation:
\begin{displaymath}
(xy) \cdot (x \alpha) \: = \: (x^2) \cdot (y \alpha)
\end{displaymath}
The module $M^{\Gamma}$ is not freely generated, and in particular
does not define a locally free sheaf on $\mbox{Spec } R^{\Gamma} 
= {\bf C}^2/\Gamma$.  However, the module $M^{\Gamma}$ does
define a reflexive rank 1 sheaf.

Thus, we see explicitly that the nontrivial orbifold Wilson line
on ${\bf C}^2/{\bf Z}_2$ describes a reflexive non-locally-free
rank 1 sheaf on ${\bf C}^2/{\bf Z}_2$.

The reflexive, non-locally-free sheaf on ${\bf C}^2/{\bf Z}_2$ can
be (non-uniquely) lifted to a sheaf ${\cal S}$ on the resolution of
${\bf C}^2/{\bf Z}_2$, such that (essentially) $(\pi_* {\cal S})^{\vee
\vee}$ is the nontrivial reflexive sheaf on ${\cal C}^2/{\bf Z}_2$.

For those readers acquainted with toric varieties,
the discussion above can be understood torically.
Describe ${\bf C}^2/{\bf Z}_2$ by a cone with edges
\begin{eqnarray*}
v_1 & = & (2,1) \\
v_2 & = & (0,1)
\end{eqnarray*}
and let $D_1$, $D_2$ denote the toric divisors corresponding
to edges $v_1$, $v_2$, respectively.
One can then check explicitly (using, for example,
methods discussed in \cite{sotvfp}) that the modules describing the
reflexive rank 1 sheaves ${\cal O}(\pm D_1)$ and ${\cal O}(\pm D_2)$
are all isomorphic (as modules, ignoring the $T$-grading),
and in particular are isomorphic to the module $M^{\Gamma}$
of invariants from the nontrivial ${\bf Z}_2$ lift, as described above.
(Note that since there is only one reflexive non-locally-free sheaf,
the sheaves ${\cal O}(+D_i)$ and ${\cal O}(-D_i)$ are necessarily
isomorphic.)

\end{document}